\documentclass{article}
\usepackage{arxiv}
\usepackage{floatrow}
\usepackage[label font=bf, 
            labelformat=simple]{subfig}
\usepackage{siunitx}
\usepackage[utf8]{inputenc} 
\usepackage[T1]{fontenc}    
\usepackage{lmodern,textcomp}
\usepackage{hyperref}       
\usepackage{url}   
\usepackage{tabularx}
\usepackage{xltabular}
\usepackage{dcolumn}
\newcolumntype{d}[1]{D{,}{,}{#1}}
\usepackage{soul}
\usepackage{booktabs}       
\usepackage{amsfonts}       
\usepackage{nicefrac}       
\usepackage{microtype}      
\usepackage{graphicx}
\usepackage{cite}
\usepackage{tabularray}
\usepackage{longtable}
\usepackage{bbding}
\usepackage{adjustbox}
\usepackage{placeins}
\usepackage{soul}
\usepackage{array}
\usepackage{amssymb, amsmath, amsthm}
\usepackage{graphicx}
\usepackage{lmodern,url}
\usepackage{makecell} 
\usepackage{cancel}
\usepackage{multirow}
\usepackage{microtype}
\usepackage{lineno}
\usepackage{xspace}
\usepackage{xcolor}
\usepackage{siunitx}
\usepackage{todonotes}
\usepackage{booktabs}
\usepackage{floatpag}
\usepackage[ruled,vlined]{algorithm2e}
\usepackage{optidef}
\usepackage{mathdots}
\usepackage{subcaption}
\usepackage{csquotes}
\usepackage{caption}

\usepackage[markup=underlined
,addedmarkup=colored
,commandnameprefix=ifneeded]{changes}

\makeatletter
\@namedef{Changes@AuthorColor}{magenta}
\colorlet{Changes@Color}{magenta}
\makeatother

\newcommand{\out}{\ensuremath{D_{d}}}

\newcommand{\dbase}{\ensuremath{D_{\text{base},d}}}

\newcommand{\vac}{\ensuremath{V_{d}}}
\newcommand{\thetavac}{\ensuremath{\theta_{V,d}}}

\newcommand{\hol}{\ensuremath{H_{d}}}
\newcommand{\thetahol}{\ensuremath{\theta_{H,d}}}

\newcommand{\tempfac}{\ensuremath{W_{d}}}
\newcommand{\amplitudetemp}{\ensuremath{\phi_{W,d}}}
\newcommand{\slopetemp}{\ensuremath{\chi_{W,d}}}
\newcommand{\shifttemp}{\ensuremath{\psi_{W,d}}}
\newcommand{\temp}{\text{tmax}}

\newcommand{\disease}{\ensuremath{C_{d}}}
\newcommand{\amplitudedisease}
{\ensuremath{\kappa_{C,d}}}
\newcommand{\slopedisease}{\ensuremath{\lambda_{C,d}}}

\graphicspath{{R1_Figures/}}

\usepackage{amsmath}

\usepackage{tikz}
\usetikzlibrary{fit}
\usetikzlibrary{positioning}
\tikzset{%
  highlight/.style={rectangle,rounded corners,fill=red!15,draw,fill opacity=0.25,thick,inner sep=0pt}
}

\title{Inferring Mobility Reductions from COVID-19 Disease Spread along the Urban-Rural Gradient}

\usepackage{authblk}

\author[1,*]{Sydney Paltra}
\author[2]{Jonas Dehning}
\author[2,3]{Viola Priesemann}
\author[1]{Kai Nagel}

\affil[1]{Technische Universität Berlin, FG Verkehrssystemplanung und Verkehrstelematik, 10623 Berlin, Germany}
\affil[2]{Max-Planck Institute for Dynamics and Self-Organization, G\"ottingen, Germany}
\affil[3]{Faculty of Physics, Georg-August-University G\"ottingen, Germany}

\affil[ ]{ {*} Corresponding author: paltra@vsp.tu-berlin.de}

\date{}

\renewcommand{\headeright}{ }
\renewcommand{\undertitle}{ }

\ifdefined \user
\advance\voffset by-0.9in
\advance\hoffset by -0.9in
\fi
\begin{document}
\ifdefined \user
\fontsize{12}{12}\selectfont
\advance\textheight by 1.8in
\fi

\maketitle
\begin{abstract}
The COVID-19 pandemic reshaped human mobility through policy interventions and voluntary behavioral changes.  
Mobility adaptions helped mitigate pandemic spread, however our knowledge which environmental, social, and demographic factors helped mobility reduction and pandemic mitigation is patchy. 
We introduce a Bayesian hierarchical model to quantify heterogeneity in mobility responses across time and space in Germany's 400 districts using anonymized mobile phone data.
Decomposing mobility into a disease-responsive component and disease-independent factors (temperature, school vacations, public holidays) allows us to quantify the impact of each factor. 
We find significant differences in reaction to disease spread along the urban-rural gradient, with large cities reducing mobility most strongly. 
Employment sectors further help explain variance in reaction strength during the first wave, while political variables gain significance during the second wave.
However, reduced mobility only partially translates to lower peak incidence, indicating the influence of other hidden factors. 
Our results identify key drivers of mobility reductions and demonstrate that mobility behavior can serve as an operational proxy for population response.
\end{abstract}


\section{Introduction} \label{sec:Introduction}

During the COVID-19 pandemic, people substantially adapted their mobility patterns as case numbers increased and official guidelines and policy evolved: 
social visits were postponed, meetings and shopping shifted online, public transport during peak-hour was avoided, stay-at-home rules were observed, and school were closed. 
These adjustments combined both voluntary adaptations due to increased personal risk and compliance to government-mandated non-pharmaceutical interventions (NPIs), making mobility an immediate indicator of population response \cite{Perra2021, Santamaria2020, Schlosser2020}. 
Indeed, investigations show that time spent away from home mediates exposure opportunities; fewer minutes spent in shared spaces translate into fewer contacts and a lower effective reproduction number\cite{Paltra2024, Perra2021}. 
Because many control measures directly affect mobility, tracking it shows whether interventions take effect, where additional support is needed, and how quickly behavior adapts. 
Importantly, these responses can be measured in near real time, due to the availability of anonymized mobile-phone data at population scale \cite{Lee2023, Mueller2021}.

The magnitude and timing of mobility reductions differ across time and space. Voluntary responses and NPI compliance depend on local information environments and social norms, workplace constraints and the feasibility of remote work, and, critically, trust in institutions \cite{Doenges2022, Lim2023, Chan2020}. 
Furthermore, structural conditions, including housing density, income, work opportunities, infrastructure, and dependency on public transport, differ along the urban-rural gradient, so the same NPIs and case numbers can yield different reductions in out-of-home duration in different locations. 
This variation matters: it changes when and where transmission accelerates, and it determines whether finite resources such as testing and hospital capacity are deployed equitably and most effectively. 
Yet most studies quantify mobility adaptations at national or international scales or, when subnational, focus on metropolitan areas \cite{Nouvellet2021, Lee2024}, leaving rural and small-town contexts comparatively underexplored \cite{Li2020, Natale2023}. 
Improving the understanding of how responses vary across the urban-rural gradient fills an essential research gap that will improve transmission modeling, targeted communication, and fair allocation of interventions \cite{Monnat2021, Assche2024}.

To quantify differences along the urban-rural gradient, we focus on Germany during the first year of the COVID-19 pandemic, quantifying changes in local mobility for all 400 German districts (\emph{Landkreise} and \emph{kreisfreie St\"adte}). 
We use out-of-home duration --- the amount of time spent away from one's home --- as our mobility measure, drawing on anonymized cellular data from a major German mobile telecommunications provider.
This sample encompasses up to 30 million devices.
Unlike reported case incidence, which serves as a key epidemiological metric but represents a delayed and composite behavioral signal influenced by testing and reporting practices, baseline transmissibility, and other protective behaviors, out-of-home duration offers a more immediate and direct measure of behavioral response to COVID-19.

In this study, we develop a Bayesian hierarchical model that decomposes out-of-home duration into a disease-responsive component and three disease-independent drivers (temperature, school vacations, and public holidays). 
The disease term combines national and local incidence through a weighted finite-memory signal and allows for pandemic fatigue, yielding a district-specific reaction strength to disease spread.
Our model successfully disentangles the different components, finding that the disease term is the leading factor impacting out-of-home duration.
We find significant difference in the reaction strength to disease spread along the urban-rural gradient, with large cities reducing most strongly.
We then relate cross-district heterogeneity in reaction strength to demographic and socioeconomic covariates.
Finally, we analyze how mobility responses interact with population density to shape peak incidence across waves. 
We find that the urban-rural differences, together with the social and economic variables, explain a large part of the variance of the reaction strength across districts. 
However, this reduction of mobility only translates partly into a reduction of the peak incidence height, as other factors remain influential. 
Overall, our work provides a deeper understanding of the relationship between the progression of the pandemic, the mobility response and how demographic, social, economic and political variables influence this relationship.

\section{Results}

\FloatBarrier
\subsection{Analysis Overview}

COVID-19 shaped the 2020 annual course of the out-of-home duration in Germany's 400 districts through a combination of government-mandated interventions and voluntary cutbacks in social and work related travel \cite{Schlosser2020, Anke2021}. Our objective is to quantify the strength of this joint effect and to identify the demographic and socioeconomic characteristics that influenced it. 
Because time spent outside the home mediates exposure opportunities, reductions in out-of-home duration strongly affect transmission dynamics and thus a wave's incidence peak \cite{Paltra2024}. 
At the same time, peak incidence reflects additional influences, including protective behaviors like mask-wearing or vaccinations and differences in the baseline reproduction numbers across districts. 
Consequently, we regress both mobility reduction and peak incidence on socioeconomic and demographic factors to understand their cross-dependencies. 
We structure the analysis in three steps:
\begin{enumerate}
\item a Bayesian model for weekly decreases in out-of-home duration driven by disease spread alongside three disease-independent factors,
\item a linear model attributing cross-district variation in these decreases to population density and demographic and socioeconomic covariates,
\item a structural equation model attributing cross-district variation in peak incidence to the same covariates and additionally to the average decrease in out-of-home duration. 
\end{enumerate}

\subsection{Impact of Disease Spread on Out-of-home Duration}
\begin{figure}[!htp]
    \centering    \includegraphics[width=\textwidth]{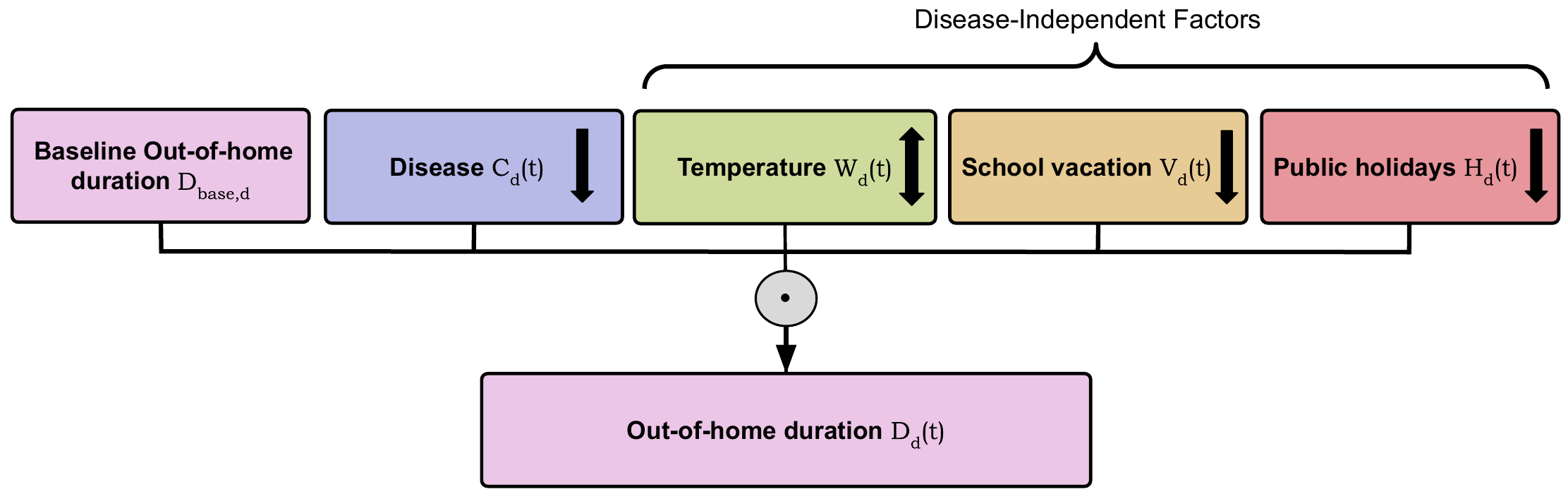}
    \caption{\textbf{Model overview illustrating the composition of the time- and district-dependent out-of-home duration.} The out-of-home duration is the product of a baseline out-of-home duration, a disease factor (based on incidence) and three disease-independent factors (temperature, school vacations, public holidays). Arrows indicate if we assume the factor to have an in- and/or decreasing effect on the out-of-home duration.}
     \label{fig:ReducedModel}
\end{figure}

Our Bayesian model spans the 52-week period from March 2020 to March 2021. 
The multiplicative model represent the out-of-home duration $\out(t)$ for each week $t \in T = \{1,\dots,52\}$ and each German district $d \in D=\{1,\dots,400\}$.
We assume that the out-of-home duration is determined by a baseline out-of-home duration, COVID-19 disease spread, and by three disease-independent factors: temperature, school vacations, and public holidays (see Fig.~\ref{fig:ReducedModel} for a reduced graphical representation of the model and Table~\ref{tab:DataSources} for introduction of underlying data bases and sources). 

\begin{figure}[!htp]
    \centering    \includegraphics[width=\textwidth]{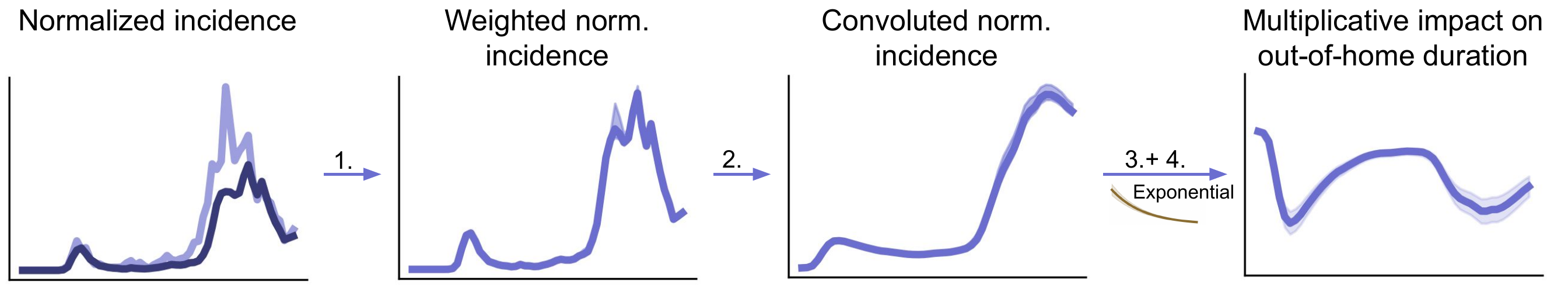}
    \caption{\textbf{Normalized local and national COVID-19 incidence are used to compute the disease factor in a four step process.} \textbf{1.} Weighted average of normalized local and national incidence is computed. \textbf{2.} Case numbers are convolved with a Gamma distribution. \textbf{3.} Multiplication of convolved case numbers with exponential decay function to integrate pandemic fatigue. \textbf{4.} Exponential of convoluted and exponentially decayed case numbers is taken.}
\label{fig:ComputationDiseaseFactor}

\end{figure}

We assume these factors are multiplicative (i.e., additive in log-space) because we assume changes in one factor do not induce changes in another, and because their effects can be expressed as multiples of a base level.
The model reads:
\begin{align}
    \out(t) = \dbase \cdot \disease(t) \cdot (\tempfac(t) \cdot \vac(t) \cdot \hol(t)). \label{eq:MainModelEquation}
\end{align}
The baseline out-of-home duration \dbase \ represents the out-of-home duration in district $d$ during a week without disease spread, without school vacations, no public holidays, and with annual average temperature.
This baseline out-of-home duration \dbase \ is multiplied by $\disease(t)$, a factor depending on the COVID-19 disease spread.
The disease factor $\disease(t)$ is based on the COVID-19 incidence rate (new weekly cases per 100,000 inhabitants) and is computed in three steps (Fig.~\ref{fig:ComputationDiseaseFactor} for illustration and Subsection~\ref{sec:MethodsDisease} for details).
\emph{First,} based on the assumption that both local and national disease dynamics influence mobility behavior, a weighted average of the normalized local and normalized national COVID-19 incidence rate is computed. 
\emph{Second,} as the population is not only influenced by the current incidence, but also by that of the recent past, incidence is convolved with a Gamma distribution.
\emph{Third,} assuming that the population's perception of risk decreases over time as pandemic fatigue sets in, the convoluted incidence is multiplied by an exponential decay function. Ergo, the same level of incidence reduces the out-of-home duration less at a later point in time.
\emph{Fourth,} assuming disease to have a multiplicative effect, we take the exponential of the convoluted and exponentially decayed incidence. This also ensures that positive incidence has a multiplicative effect smaller than one.

The product of the baseline out-of-home duration \dbase \  and disease factor $\disease(t)$ is further multiplied by three disease-independent factors:
\begin{itemize}
\item Temperature factor $\tempfac(t)$. When temperatures are high, this factor increases the out-of-home~duration, while when temperatures are low, this factor decreases the out-of-home~duration \cite{Jaerv2014, Schoenfelder2016}. 
\item School vacation factor $\vac(t)$. School vacations decrease the weekly out-of-home~duration, as children do not attend school and parents take time off work to look after their children \cite{Lai2022}.
\item Public holiday factor $\hol(t)$. Most employees are not required to work on public holidays, leading to a decrease in out-of-home duration \cite{Cool2007, Wheeler2010, Lai2022}.
\end{itemize}
The strength of each of these factors is allowed to vary between districts, we only assume that the strength is to some degree similar across districts (hierarchical modeling). 
We focus on the period from March 2020 to March 2021 because COVID-19 vaccines were not yet widely available during this time frame, thus ensuring that vaccinations did not affect the out-of-home duration.
We perform Bayesian inference for the parameters of our model using Markov-Chain Monte Carlo (MCMC) sampling (see Section~\ref{sec:methods:inference} for details).

\begin{table}
    \centering
    \begin{tabular}{|l|p{10cm}l|}
    \hline
    \textbf{Factor} & \textbf{Based on} & \textbf{Source} \\ \hline
    $\out(t)$ & Based on cell-based mobility data. For each zip code, the out-of-home duration sum over all recorded persons (in seconds) is available. Zip codes are grouped according to their district. For each district and each day, the 
    out-of-home duration sum (in seconds) is computed, divided by the number of people living in this district and by 3600 to obtain the out-of-home duration (in hours) of an average person in the $d$. From the daily out-of-home durations, the weekly average is computed \cite{Mueller2024}.& \cite{Balmer2022} \\
    $\disease(t)$ & COVID-19 7-Day Incidence rate for each district $d$ and national COVID-19 7-Day Incidence rate. & RKI \cite{rki2023} \\
    $\tempfac(t)$ & Weekly average of daily maximum temperature (in $^\circ C$) for district $d$. If daily maximum temperature was unavailable any time during the study period for district $d$, we considered the average daily maximum temperature of the neighboring districts. & Meteostat \cite{Meteostat2023} \\
    $\vac(t)$     & School vacations differ on federal state, not district level. Sum of school vacation days for week $t$. Children in Germany go to school Monday--Friday; maximum number of school vacation days in week $t$ is 5. & Schulferien.org\cite{Schulferien2025}\\
    $\hol(t)$   & Public holidays differ on federal state, not district level. Sum of public holidays that fall on weekday in week $t$. & Schulferien.org\cite{Schulferien2025}\\ \hline
    \end{tabular}
    \caption{\textbf{Data bases for dependent and independent variables.} Exploratory analysis and conception of factors may be found in in Section~\nameref{sec:Methods}.}
    \label{tab:DataSources}
\end{table}

\begin{figure}[!htp]
    \centering
\includegraphics[width=\textwidth]{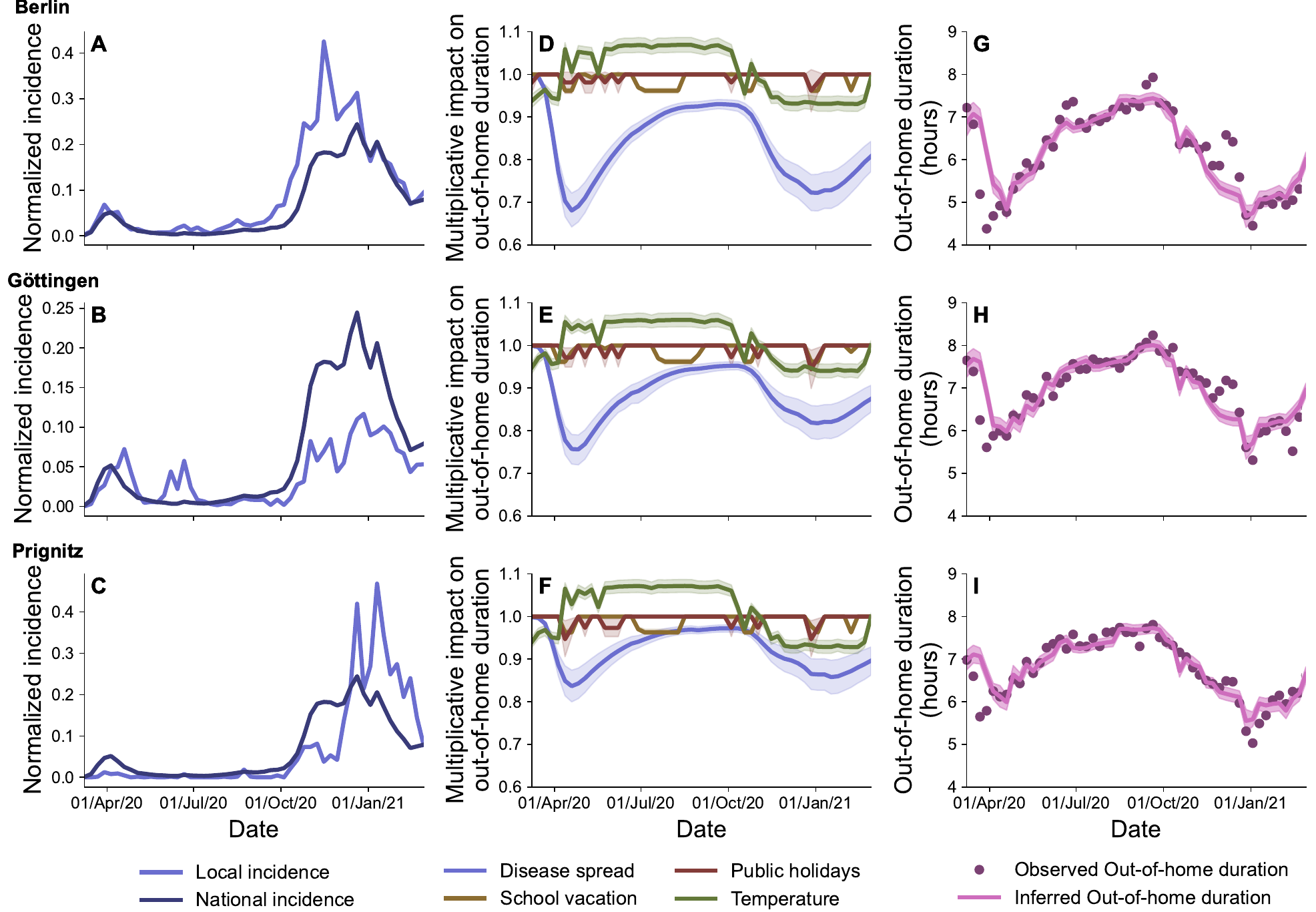}
    \caption{\textbf{Using Bayesian inference we successfully infer the annual course of the out-of-home duration for the three exemplary districts Berlin, Göttingen, and Prignitz.} The authors chose to depict their home districts of Berlin and Göttingen, together with Prignitz, the district with the lowest population density.
    In the middle and the right column, lines represent median factors and the ribbons represent 95\% Bayesian credibility intervals.
    \textbf{A-C.} Normalized local and national incidences. All three exemplary districts experienced a first (spring 2020) and second (winter 2020/2021) COVID-19 wave. 
    \textbf{D-F.} Multiplicative impact of the disease factor $\disease(t)$ and the three disease-independent factors $\tempfac(t), \vac(t)$, and $\hol(t)$. The disease factor $\disease(t)$ decreases the out-of-home duration most strongly, followed by smaller effects of $\tempfac(t)$, and only modest reductions due to $\vac(t)$ and $\hol(t)$.
    \textbf{G-I.} Observed vs inferred out-of-home duration. The model successfully infers the annual course of the out-of-home duration, specifically the shape of the two major decreases in spring 2020 and winter 2020/2021.}
\label{fig:MultImpactsBerlin}
\end{figure}

Our model~\eqref{eq:MainModelEquation} successfully fits the observed out-of-home duration between March~2020 and March~2021 (Fig.~\ref{fig:MultImpactsBerlin} right column for exemplary depictions of Berlin, Göttingen, and Prignitz). 
The model infers the overall shape of the two major decreases in out-of-home duration during the first year of the COVID-19 pandemic: the sharp decline in spring 2020 during the initial COVID-19 wave and the second decline in winter 2020/2021 during the second COVID-19 wave.
While the model successfully captures these trends, it leaves opportunity for improvement at two specific time points.
First, the model predicts the local minimal out-of-home duration in spring 2020 too late (Model: April 5th, 2020 vs Data: March 15th, 2020).
Second, the model does not capture a temporary spike in out-of-home duration observed in late November~2020 to early December~2020, which interrupted the otherwise declining trend in out-of-home duration.
A potential explanation for this spike is the looming stricter lockdown that was finally announced on December 13th,~2020 and took effect on December 16th,~2020.
This encouraged advanced Christmas shopping and increased activity.
But overall, the model fits most variance well, including some short-term variations explained by school vacations and public holidays.

The disease factor \disease \ decreased the out-of-home duration substantially more in spring~2020 than in winter~2020/2021. 
All three disease-independent factors, school vacations, public holidays, and temperature, significantly influence the out-of-home duration during the two COVID-19 waves. 
Temperature has a decreasing effect only until mid-April~2020 and an increasing effect over the warm summer months (May -- September). 
Overall, the findings demonstrate that disease spread dominated the out-of-home duration in districts as diverse as Berlin, Göttingen, and Prignitz, while  seasonal and holiday factors provided secondary but consistent reinforcing effects.

In the three exemplary districts of Berlin, Göttingen, and Prignitz, the disease spread had a stronger multiplicative effect on out-of-home duration than the three disease-independent factors (Fig.\ref{fig:MultImpactsBerlin}~middle column), with substantially larger reductions in spring~2020 than in winter~2020/2021. 
The three disease-independent factors -- school vacations, public holidays, and temperature -- significantly influenced out-of-home duration, provide secondary but consistent reinforcing effects, with temperature decreasing mobility until mid-April~2020 and increasing it during the warm summer months (May--September).

\begin{figure}[!htp]
    \centering
\includegraphics[width=\textwidth]{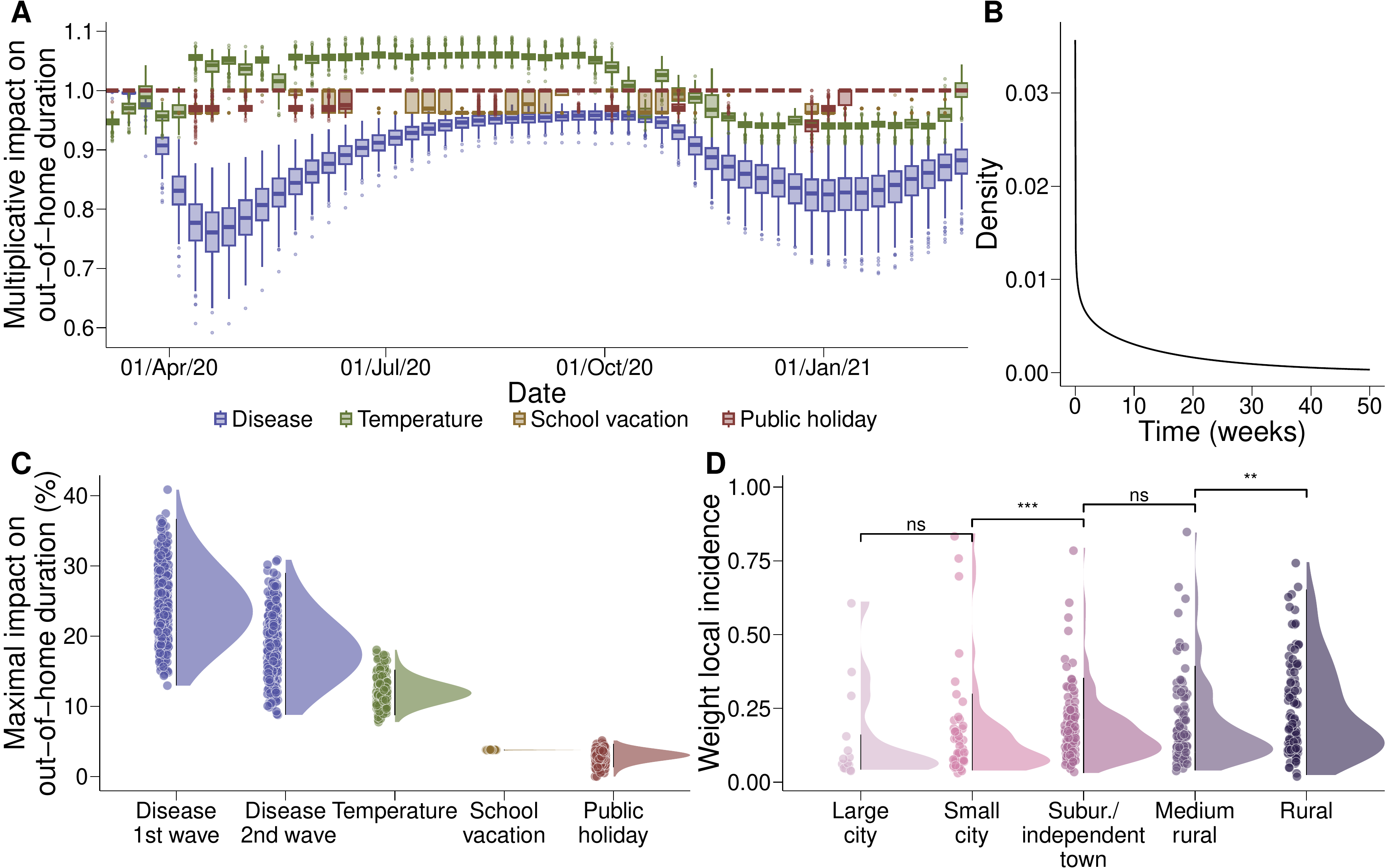}
\caption{\textbf{A. Over time, the decreasing effect of the disease factor relaxes.}
Distribution of multiplicative impact across districts, using the median multiplicative impact for each district. Across districts, the disease factor emerges as the strongest influence, with pandemic fatigue evident in every district. Due to temperature fluctuations, out-of-home duration is approximately 10\% higher in summer than in winter. School vacations and public holidays lead only to modest reductions.
\textbf{B. Gamma distribution delays the effect of the weighted normalized incidence.} The weighted norm. incidence is convoluted with a gamma distribution to represent that the out-of-home duration in week $t$ is not only instantaneously influenced by disease spread in week $t$, but also by disease spread in the recent past. 
\textbf{C. The disease factor emerges as the strongest influence on out-of-home duration.} Distribution of effect sizes, in other words size of impact on out-of-home duration, across districts, using the median effect size for each district. While the effect size of the disease factor decreases from the first to the second wave, it still influences out-of-home duration most strongly. Secondary effects due to temperature, followed by only modest effects due to school vacation and public holiday.
\textbf{D. Local incidence carries only a minority of the weight in explaining mobility behavior.} Distribution of weight of normalized local incidence in disease factor $\disease(t)$. Across districts, the majority of weight in explaining local mobility behavior is placed on national rather than local incidence, aligning with the introduction of nation-wide NPIs and communication strategies.
Stars indicate t-test determining difference in mean across district types to be statistically significant, with significance levels of  $^{\star \star \star}$ representing $p~<~0.001$, $^{\star \star}$ representing $p~<~0.01$, and ns representing $p>0.05$.}
\label{fig:DistributionMultImpacts}
\end{figure}

Expanding the analysis from the three exemplary districts to all 400 German districts (Fig.~\ref{fig:DistributionMultImpacts}~A), the disease factor consistently emerges as the strongest influence on out-of-home duration, confirming the pattern observed for Berlin, Göttingen, and Prignitz:
During the first wave, the disease factor $\disease(t)$ reduces the out-of-home duration most strongly in the week ending April 19th,~2020, reaching a median multiplicative impact of 0.77, IQR: ([0.73, 0.80]) (Fig.~\ref{fig:DistributionMultImpacts}~C).
Despite higher incidences during the winter 2020/2021 wave, the disease factor reduces out-of-home duration less than during spring 2020, reaching a median maximal multiplicative impact of 0.82 (IQR: [0.79, 0.85]). 
This suggests that pandemic fatigue fully compensates the effect of higher incidences.
The difference in out-of-home duration between winter and summer is about 11\%,
with a median multiplicative impact of 1.06 (IQR: [1.05, 1.07]) during the warm summer months (May--September) and 0.95 (IQR: [0.94, 0.95]) during the winter.
The decreasing impact of school vacations and public holidays is even smaller: 
The median difference between weeks with school vacations and weeks without is 4\% (same effect is inferred for all districts; boxes in Fig.~\ref{fig:DistributionMultImpacts}~C represent that during the corresponding week school vacations occurred in some districts but not in others, rather than a difference in effect size). 
The median difference between weeks with public holiday(s) and without is 3\%.
In summary, across districts and just like for the exemplary districts of Berlin, Göttingen, and Prignitz, disease spread most strongly reduces the out-of-home duration, followed by the influence of temperature and modest reductions due to school vacations and public holidays.

\FloatBarrier

\FloatBarrier
\subsection{Stronger Mobility Reductions in More Densely-Populated Districts} \label{sec:ReacStrengthDistType}
\begin{figure}[!htp]
    \centering
\includegraphics[width=\textwidth]{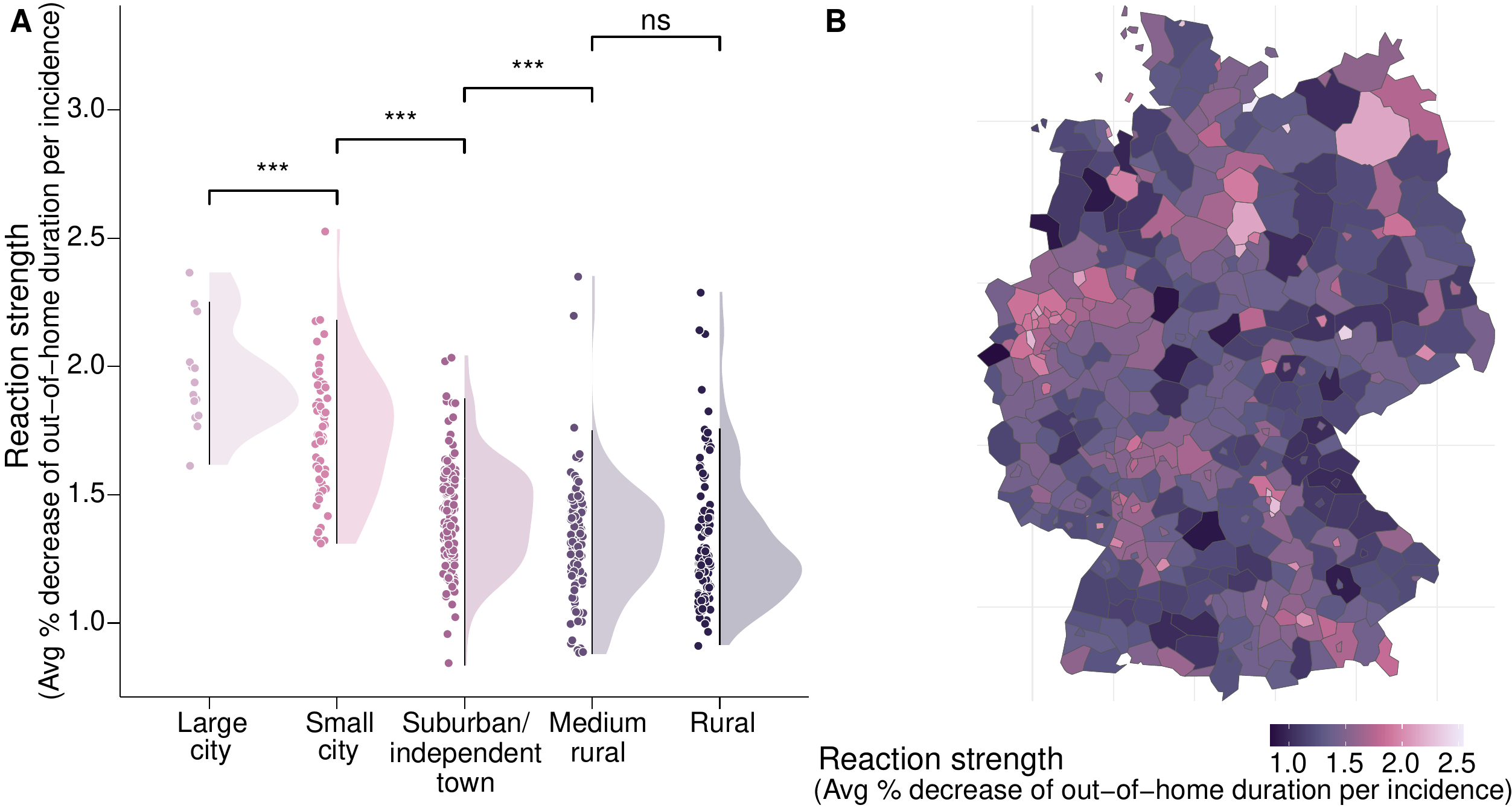}
\caption{\textbf{Differentiation by district type and spatial depiction strongest reaction strength in cities and parts of the west of Germany.} In either panel, incidence represents new weekly cases per 100,000 inhabitants. \textbf{A.} Large cities display the greatest average reaction strength, with reaction strength structurally decreasing along the urban-rural gradient. Stars indicate t-test determining difference in mean across district types to be statistically significant, with significance levels of  $^{\star \star \star}$ representing $ p~<~0.001$ and ns representing $p>0.05$. \textbf{B.} Depicting the reaction strength spatially highlights great reaction strength in cities (small polygons) and in the west of Germany, contrasting with modest reaction strengths in parts of the former GDR and northern Bavaria.}
\label{fig:FirstModelResults}
\end{figure}

\vspace{-0.4em} 
In order to understand the differences in reduction of out-of-home duration between districts, we first investigate how much it depends on rural versus urban differences. 
To this end, we classified districts into five district types: large city, small city, suburban/independent town, medium rural, and rural (using the classification system of the \emph{German Federal Institute for Research on Building, Urban Affairs, and Spatial Development}).
We do not directly use the reduction of out-of-home duration as our variable of interest, but the \emph{reaction strength} of the out-of-home duration with respect to the incidence. 
We use reaction strength as we expect that for districts with high local incidence, the decrease in out-of-home duration will be stronger (even if that effect is not so strong, as most of the district are more strongly tied to the national incidence, Fig.~\ref{fig:DistributionMultImpacts}~B).
As reaction strength we used the average impact of the incidence on the reduction of out-of-home duration, which is mathematically the integral of our exponential decay function.
Comparison across district types revealed a large differences: large cities exhibited on average the largest reaction strength, while rural districts showed the weakest response (Fig.~\ref{fig:FirstModelResults}~A).
T-tests comparing mean reaction strength across district types confirmed significant differences between the four more densely populated district types:  large city vs small city ($p < 0.01$),  small city vs suburban/independent town ($p<0.001$), and suburban/independent town vs medium rural ($p<0.001$).
Finally, the difference in mean reaction strength reaches a plateau, with no statistically significant difference observed between medium rural and rural districts ($p > 0.05$).
Overall, this demonstrates a clear urban-rural gradient in response to disease spread, with reaction strength decreasing systematically from large cities to rural areas.

However, differentiation by district type does not suffice to fully explain the observed differences in mobility reduction in the 400 districts (see Fig.~\ref{fig:FirstModelResults}~B, in general smaller polygons represent cities).
For instance, we observe strong reactions in the west of Germany, particularly all across North Rhine-Westphalia.
Meanwhile, northern Bavaria (specifically the Upper Palatinate region) shows only moderate reactions, even in the city districts. 
Furthermore, we are interested in whether urbanness 
is really the main factor that influences urban-rural differences in mobility reduction, or whether partially-collinear demographic or economic factors also play a role. 

To better understand which factors are responsible for the observed differences in mobility reduction, we regressed nineteen potentially relevant demographic, socioeconomic, and political variables that encode differences between districts on the \emph{reaction strength during the first wave} and on the \emph{reaction strength during second wave} separately (defined as the integral over the corresponding wave, see Section~\ref{sec:RegressionReactionStrength} for details). 
More precisely, as demographic variables we considered population density, average age, and share of 65+ year old people; as economic variables we considered the average income, employment and unemployment rate, economic sectors (agriculture, forestry fisheries, manufacturing sector, construction, service sectors, trace, transport, hospitality, information, communication (TTHIC) sectors, finance sector); and as sociological and political variables we considered the share of small children in childcare, the voter turnout, and share of votes of any of the political parties that received more than 5\% of votes in the 2021 federal election (see Supplementary Table~\ref{tab:RegressionVariableOverview} for overview and summary statistics).
Exhaustive search and comparison of model selection criteria (adjusted $R^2$, Mallow's $C_p$, AIC, and BIC) determined which explanatory variables were considered in the final regression on reaction strength. 
In the regression, we then used the set of variables for both waves that are significant during either wave (see Subsections~\ref{sec:RecStrengthBestSubset} and \ref{sec:RecStrengthModelSelection} for details). 

\begin{figure}[!htp]
    \centering    \includegraphics[width=\textwidth]{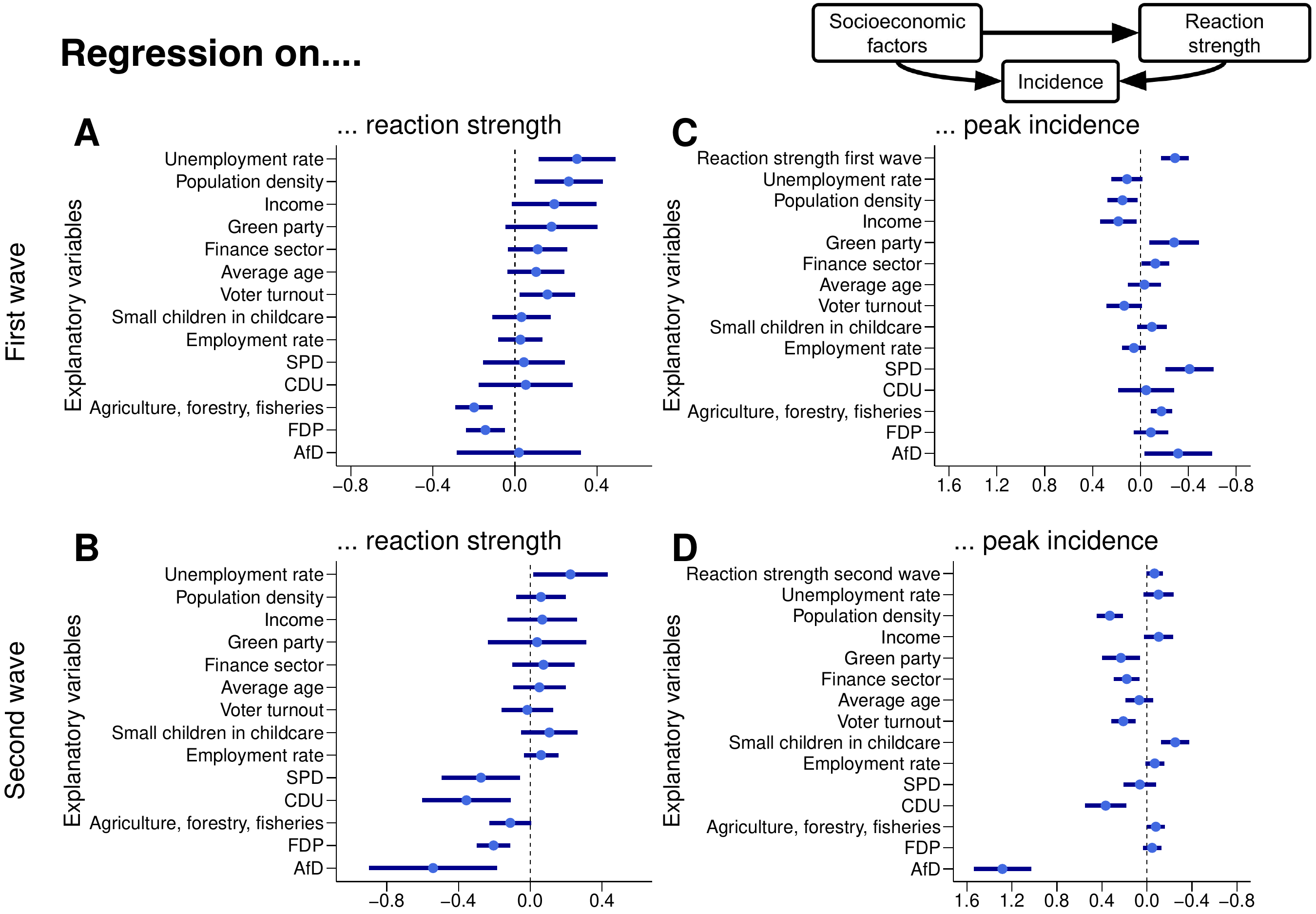}     \caption{\textbf{Multiple regressions and structural equation models highlight the complex interrelation of reaction strength and peak incidence.} 
    \textbf{A,B.} 
    Regression of demographic, socioeconomic, and political variables on \emph{reaction strength first wave} (A) and on \emph{reaction strength second wave} (B), reveals a shift in explanatory factors from economic and demographic factors to political conviction.
    \textbf{C,D.} Regression of reaction strength and the same demographic, socioeconomic, and political variables  on \emph{peak incidence first wave} (C) and \emph{peak incidence second wave} (D), highlights that peak height depends, but not exclusively, on mobility reductions.
    }
\label{fig:Regression_Contributions}
\end{figure}

We find that during the first COVID-19 wave, \emph{unemployment rate} and \emph{population density} contribute most strongly to explaining variance in reaction strength, with similar strong effects (Figure~\ref{fig:Regression_Contributions}~A and Table~\ref{tab:RegressionOutputReacStrength} for detailed model output). 
The strong contribution of \emph{population density} was expected from the large differences in reaction strength between rural and urban areas (Figure~\ref{fig:FirstModelResults}~A). 
The strong contribution of \emph{unemployment rate} may be explained by the fact that the economy of high-unemployment districts is often concentrated in sectors that are more vulnerable to pandemic disruptions (retail, hospitality, manual labor).
During an outbreak, these jobs get shut down or drastically reduced, yielding a stronger reaction \cite{Boehme2020}. 
Another explanation may be that high-unemployment districts often have a generally weak economy such that even without strict lockdowns, economic hardship enforces staying home.
Furthermore, the regression coefficients of three other independent variables are also significantly different from zero ($p<0.05$, not multiple comparison corrected): \emph{agriculture, forestry, and fisheries}, \emph{voter turnout}, and \emph{FDP}.
For \emph{income}, the regression coefficient is significantly different from zero at a significance level of $p=0.1$.
We have some proposals for the causal relationship between the decrease in out-of-home duration and these four variables:
\emph{Agriculture, forestry, fishers} is an economic sectors with little to no home office feasibility, forcing employees into the workplace, thus limiting \emph{reaction strength}.
\emph{Voter turnout} may be understood as a measure of political trust, and as such, high voter turnout enhances adherence to guidelines and consequently staying at home \cite{Brodeur2020, Shanka2022}.
Voting for \emph{FDP} correlates with political convictions that may discourage reducing one's mobility.
A higher average \emph{income} is correlated with white-collar jobs, which allow for remote work more readily, increasing reaction strength \cite{Dingel2020}. 
To conclude, however, these variables might not have direct causal connections, and may rather be indicative of underlying social and economic conditions that shape mobility responses.

Regressing on \emph{reaction strength second wave} reveals a shift in explanatory power of independent variables, more concretely, away from labor and economic variables and towards political variables.
The regression coefficients of four political party variables are now negative and significantly different from zero (significance level of $p=0.05)$: \emph{CDU}, \emph{SPD}, \emph{FDP}, and \emph{AfD} (Fig.~\ref{fig:Regression_Contributions}~C and Table~\ref{tab:RegressionOutputReacStrength} for detailed model output).
A possible explanation may be that during the first wave, nonpharmaceutical interventions and communication strategies were implemented nationally, uniting society in an attempt to flatten the curve.
These national implementations dictated home office feasibility and on-work site necessity clearly.
During the second wave, this authority shifted more and more from the national to the federal state level, highlighting the different approaches of the political parties (and consequently their voters) to disease spread.
The regression coefficient of \emph{unemployment rate} is still statistically significant from zero ($p < 0.05)$, but  comparable in magnitude to those of the political variables.
In contrast, the regression coefficient of \emph{population density} is not statistically different from zero anymore. 

Together, our findings suggest that mobility responses shifted from being primarily driven by structural economic and demographic factors during the first wave to being increasingly shaped by election results and regional governance during the second wave.

\FloatBarrier
\subsection{Cross-district Differences in Peak Incidence}

Relevant for pandemic control, however, is not the strength of mobility reductions, but rather the peak height of the infection waves, which determines the maximal burden on the health care system. 
Peak height, in turn, depends on the decrease of mobility, but not exclusively. 
There certainly is a causal relationship, as a reduction of out-of-home duration limits contacts and consequently reduces infection opportunities and thus the effective reproduction number and peak height \cite{Nouvellet2021, Badr2020}. 
However, other factors also play a role, such as the local baseline reproduction number. 

\begin{figure}[!htp]
    \centering    \includegraphics[width=\textwidth]{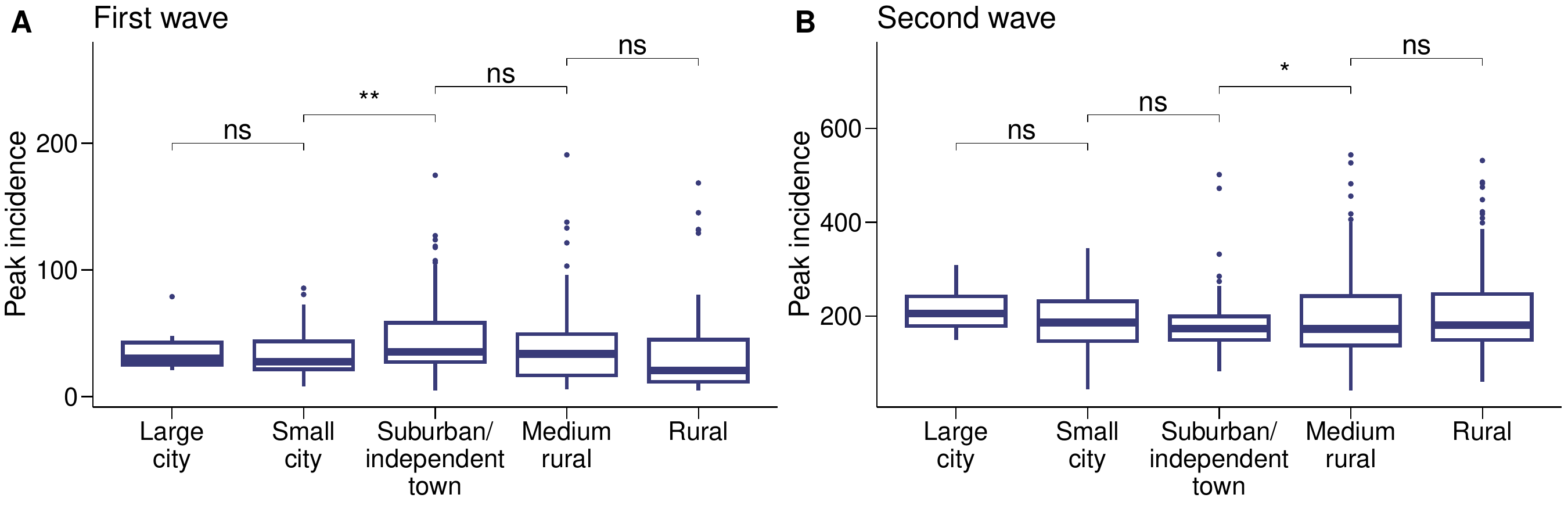}    \caption{\textbf{Stronger reaction strength does not necessarily yield smaller peak height.} In both panels, peak incidence represents the 90th percentile of incidence during the corresponding wave. \textbf{A.} For the first wave, cities demonstrate smaller peak heights than suburban/independent towns, but so do the rural areas. \textbf{B.} For the second wave, peak incidence decreases along the urban-rural gradient until it levels off for medium rural and rural districts. In both panels, stars indicate that the t-test determined the means across district types to be statistically significant, with significance levels of $^{\star \star}$ representing p<0.01, $^{\star}$ representing p<0.05 and ns representing p>0.05.}
    \label{fig:PeakIncidence}
\end{figure}

A key indication is that during the first wave the peak height does not reveal an urban-rural gradient which we might have expected given the stronger decrease of mobility in urban districts (Figure~\ref{fig:PeakIncidence}~A). 
Moreover, during the second wave, even though we observed the strongest reactions in cities, they also demonstrated the largest peak height (Figure~\ref{fig:PeakIncidence}~B).
The lack of urban-rural gradient suggests that rural areas may have a smaller baseline reproduction number.
Differences in baseline reproduction may then, in turn, be traced back to demographic and socioeconomic factors.

To investigate these hypotheses, we used structural equation modeling, regressing on \emph{peak incidence} of the first and the second wave separately. 
As explanatory variables we considered the demographic, socioeconomic, and political variables of the previous section.
\emph{Reaction strength} entered the structural equation models as a mediator.
If the peak incidence of either wave was driven solely by mobility reductions, only reaction strength should be significant, with other variables’ effects mediated through it.

For both waves the reaction strength, i.e.~our mobility measure, contributes to a decrease in peak incidence (Figure~\ref{fig:Regression_Contributions}~C,~D, for the second wave, only at a significance level of $p<0.1$. Note that the x-axis is flipped, for easier comparison with panels A and B). 
In contrast, population density has an increasing effect, counteracting the reaction strength, which might explain nearly the constant or increasing peak height from rural to urban districts. 
Indeed, urban areas must react more strongly in order to compensate for their higher population density, which often coincides with crowded public spaces, smaller living spaces, and a higher share of public transport, which all facilitate disease spread \cite{Alirol2011, Teller2021}. 
Otherwise, $\emph{agriculture, forestry, fisheries}$ is the only variable whose regression coefficient does not change its sign between the first and second wave. 
Thus, interpreting the influence of the other socioeconomic and political factors is difficult, despite them having a relatively high impact during both waves. 
Similarly as for the regression on reaction strength, this might be a indication that the context of the first and second wave was different, from mitigation being driven more by Germany-wide interventions and restrictions during the first wave, to more federal state wide interventions and voluntary cautious behavior during the second wave. 
Taken together, these results show that our measure of mobility is only one of the factors determining the COVID-19 spread 
and spreading models and predictions may be improved by taking into account the composite of mobility and other environmental factors.

\FloatBarrier
\section{Discussion}
We quantified the change in local mobility in Germany's 400 districts during the first year of the COVID-19 pandemic year using out-of-home durations derived from anonymized cellular data. 
For each week, our hierarchical Bayesian model decomposes the out-of-home duration into a product of a baseline out-of-home duration, a disease factor based on the local and national COVID-19 incidence, and three disease-independent factors (temperature, school vacations, public holidays). 
Three findings stand out: 
First, the model successfully reproduces both the timing and the magnitude of the major reductions in out-of-home duration across districts. This required incorporating a disease factor with diminishing impact over time, reflecting pandemic fatigue. 
Second, national incidence carries more weight in explaining mobility behavior than local incidence.
Third, districts' reductions in out-of-home duration in response to disease spread varied significantly across space, decreasing systematically from large cities to rural areas.

Our Bayesian inference model attributes most of the reduction in out-of-home duration to disease spread, secondary effects to temperature  and only modest reduction to calendar effects (school vacations, public holidays), aligning with evidence that policy and perceived risk jointly drive mobility changes \cite{Nouvellet2021, Badr2020, Murray2021}. 
Our framework, however, cannot fully disentangle the drivers behind the impact of disease spread on behavior. 
The disease factor, which captures the reaction to incidence, combines government-mandated and voluntary responses that co-vary with incidence and media coverage \cite{Yan2021, Summan2021}, which must be considered when interpreting our results.
Importantly, to infer the out-of-home duration successfully, the disease factor has to incorporate that the reduction in out-of-home duration resulting from disease spread weakens over time.
This gradual relaxation is consistent with survey-based evidence on declining adherence \cite{Petherick2021, world2020pandemic}, though the underlying mechanisms --- habituation, economic pressure, poor communication --- need further investigation and may vary between subpopulations.

The stronger influence of national compared to local incidence suggests that information and policy signals were integrated on a national level. 
This is plausible in Germany, where nationwide announcements and broadly coordinated measures dominated in 2020 \cite{Hale2021}. 
Heavily weighting the national incidence also reflects that mobility behaviors are coupled across districts through commuting and media consumption. 
However, we do not claim this weight to be universal: in settings with highly decentralized policies or localized outbreaks, local incidence may matter more. 
 Understanding how people weight varying signals remains an open problem with implications for the timing and targeting of NPIs and communication strategies.

We found districts to vary significantly in how much they reduced out-of-home duration in response to disease spread, with reductions decreasing systematically from large cities to rural areas.
The observed urban-rural gradient in out-of-home duration reduction can be motivated by higher vulnerability and baseline contact opportunities in cities \cite{Monnat2021, Assche2024}, yet density alone cannot fully account for the variation.
Regional clusters indicate that inter-district connectivity, sectoral composition, and early outbreak histories also matter \cite{Schlosser2020}. 
Political and social context likely contributes as well: our trust-proxy (voter turnout) correlates with stronger reductions, echoing work linking trust in government and institutions to adherence \cite{Petherick2021, Lim2023}. 
Differences in home office feasibility by occupation further support larger potential reductions in higher-income, service-oriented urban districts \cite{Dingel2020}. 
These observations are supported by our regression, where population density and unemployment rate were the dominant correlates, followed by modest contributions of additional socio-economic and political covariates (voter turnout, income, industries with little home office feasibility, voting shares). 

Importantly, however, stronger mobility reductions did not automatically translate into smaller local incidence peak height: 
We did not observe a clear urban-rural gradient in peak height for the first wave, and cities experienced larger peaks during the second wave despite stronger reductions. 

There are three caveats one has to keep in mind when interpreting our results. 
First, smartphone-based measures can under-represent specific demographics (e.g., the very old or young) and essential workers, potentially biasing urban-rural comparisons \cite{Wesolowski2016}. Second, district-level covariates risk ecological fallacy. 
As coefficients are estimated on between-district variation, their magnitudes also reflect how much each covariate varies across districts. Variables with limited between-district variation (even if heterogeneous within districts) will yield smaller standardized regression coefficients.
Individual-level data could improve model accuracy and help understand equity implications.
Third, the inferred importance of population density is robust and aligns with known structural differences, but the socio-economic factors with smaller coefficients -- income, unemployment, childcare, voter turnout -- may be proxies for latent constructs (job mix, employer practices, information environments). 
Natural experiments that shift these factors, for instance change in home office feasibility or closure of childcare facilities, would be needed to isolate mechanisms \cite{Yan2021, Summan2021}.
As a result, our regression and structural equation model should be read as explanatory rather than causal.

A central methodological choice we made is the mobility measure.
We argue that the out-of-home duration is an appropriate choice, as it directly tracks time potentially spent in shared physical spaces and is comparable across districts and weeks. 
This contrasts with distance-based measures, trip counts, or venue visitation indices, which capture different behavioral facets \cite{Grantz2020, Oliver2020, Buckee2020}. 
Out-of-home duration aggregates across trip purposes without requiring venue classification and maps naturally to contact time in many settings, though it does not distinguish crowded from sparse environments, indoor from outdoor time, or clustering at specific locations --- all of which influence transmission dynamics\cite{Wesolowski2016, Chang2021, Rader2020}. 
This limitation becomes evident in the imperfect link between reductions in out-of-home duration and peak incidence: Within-own household transmissions, crowding in essential workplaces, and lack of other self-protective behaviors (e.g. mask-wearing) can sustain spread even when time away from home declines \cite{Madewell2020}. 
Additionally, cities may need larger reductions to offset inherently higher contact opportunities linked to crowding  in places ranging from home to public transport vehicles \cite{Rader2020}. 
This underscores the need to integrate a mobility quantity with context (who, where, how crowded) to improve predictive value. 
Combining duration with occupancy and indoor-air proxies, and modeling network structure explicitly \cite{Schlosser2020, Chang2021}, are promising directions.

Some aspects of our modeling approach present opportunities for improvement. 
For instance, our disease factor combines local and national incidence for Germany but does not include cross-border epidemiological dynamics, which may affect border regions \cite{Docquier2022}. 
Furthermore, we observe that minimal out-of-home duration in early March~2020 preceded incidence peak; incorporating additional drivers --- such as media and risk-perception signals or international incidence --- may improve timing, especially at the start of waves \cite{Petherick2021, world2020pandemic, Grantz2020, Goolsbee2021}.

Despite these limitations, our findings have practical implications.
As out-of-home duration is measurable in near real time, it can serve as an operational proxy for population response, complementing slower epidemiologic indicators \cite{Grantz2020}. 
Further, the observed urban-rural gradient suggests calibrating communication and support locally: dense cities may require earlier or stronger measures to achieve comparable risk reduction, while rural areas may benefit from tailored strategies that address different job mixes and transport patterns \cite{Monnat2021, Assche2024}.

\section*{Author Contributions Statement}

Conceptualization: SP, JD, VP, KN \\
Data curation: SP  \\
Formal analysis: SP \\
Funding acquisition: VP, KN \\
Investigation: SP \\
Methodology: SP, JD \\
Project administration: SP, VP, KN \\
Resources: VP, KN \\
Software: SP, JD \\
Supervision: JD, VP, KN \\	
Validation: all \\
Visualization: SP \\	
Writing - Original Draft: SP, JD \\	
Writing - Review \& Editing: all 

\section*{Acknowledgements}

We thank Emil Iftekhar for his contribution to an early model version and Arne Gottwald and Piklu Mallick for the valuable comments they provided. This work was funded by the German Federal Ministry for Education and Research, namely the infoXpand project (031L0300A, 031L0300D) and the MODUS-Covid project (031L0302A), and TU Berlin. Claude AI and ChatGPT were used for grammar checks in the main text.

\section*{Competing Interests Statement}

The authors declare no competing interests.

\renewcommand{\refname}{References}

\bibliographystyle{ieeetr}
\bibliography{main}

\section{Methods} \label{sec:Methods}

\subsection{Input Variables}

\FloatBarrier
\subsubsection{Temperature}

\begin{figure}[!htp]
    \centering    \includegraphics[width=0.5\textwidth]{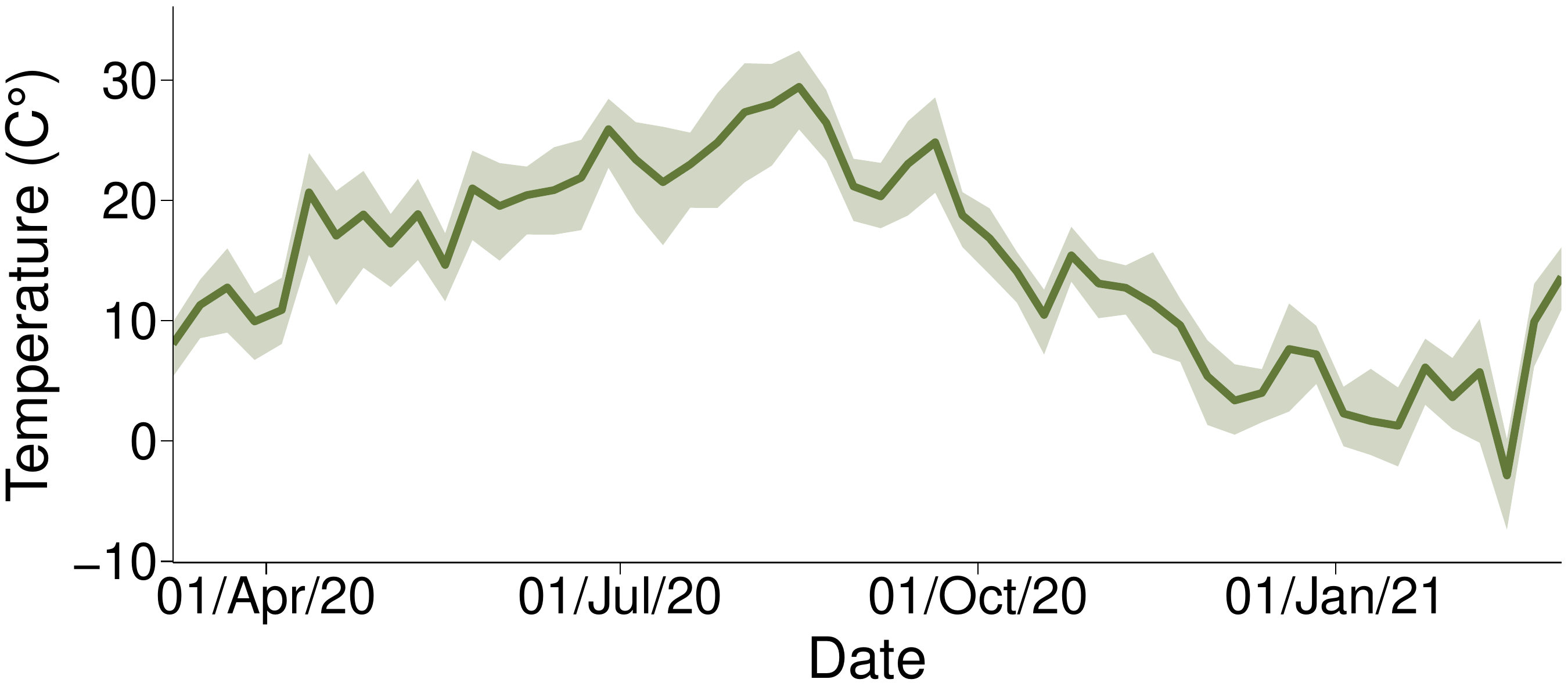}
    \caption{\textbf{Temperature follows a seasonal pattern, with highs in summer and lows in winter.} For each district, the weekly average of the daily maximum temperature is used. The line represents the mean across districts, ribbons represent the 95\% interpercentile range. The global maximum temperature occurs mid-August 2020. From then onwards, temperature steadily decreases until it reaches a global minimum in February 2021.}
    \label{fig:InputTemp}
\end{figure}

In Germany, weather influences the out-of-home duration throughout the year, with warmer temperatures motivating individuals to spend extended periods outside their homes, whereas lower temperatures promote staying at home \cite{Jaerv2014, Schoenfelder2016}.
Temperatures hereby follow a seasonal pattern: 
During the winter months --- November to February --- temperatures are low, partly dropping below the freezing point, while the summer months --- June to August --- bring high temperatures up to $30^\circ$C (Fig.~\ref{fig:InputTemp}).
During the study period, the maximal average german temperature was recorded mid-August~2020, whereas the minimal average temperature was recorded in February~2021.
For our analysis, we consider the weekly average of the daily maximum temperature. 
Whenever temperature data is unavailable for a district, we consider the average temperature of all bordering districts for which temperature data is available.
In sum, weather affects the out-of-home duration, with higher temperatures encouraging time spent outside one's home and lower temperatures promoting staying indoors, following Germany's seasonal pattern of winter lows (below freezing point) and summer highs (up to $30^\circ$C).

\subsubsection{School Vacation}

\begin{figure}[!htp]
    \centering    \includegraphics[width=0.5\textwidth]{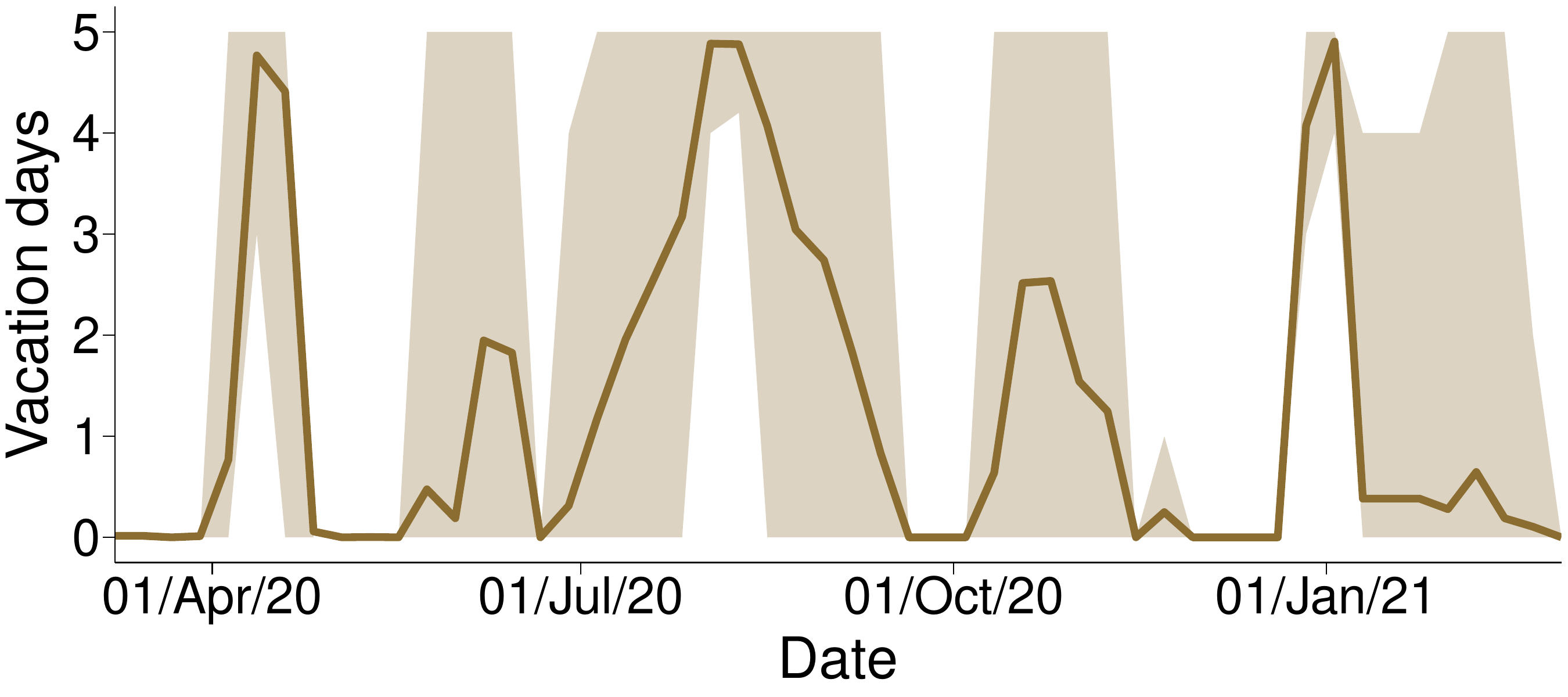}
    \caption{\textbf{Summer break is the most extensive vacation period in Germany, shorter periods in the spring and winter.} For each week, the number of school vacation days (Monday -- Friday) is summed up.
    Line represents the mean across the 16 federal states, ribbons represent the 95\% interpercentile range. Most federal states have four breaks: spring, summer, fall, and winter, with summer vacation (July -- September) representing the longest break period.}
    \label{fig:InputSchools}
\end{figure}

School vacations decrease the out-of-home duration through two mechanisms: 
directly by keeping children from going to school and indirectly by requiring parents to take vacation days or work remotely to look after their children.
In Germany, school vacations differ by federal state, with most states having spring, summer, fall, and winter breaks.
For each week, we sum up the number of school vacation days (Fig.~\ref{fig:InputSchools}). 
As children in Germany go to school from Monday until Friday, the maximum number of school vacation days per week is five. 
Summer vacation (between July and September), with a duration of six weeks in most federal states, represents the longest break period.
Overall, school vacations decrease the weekly out-of-home duration by keeping children home from school and prompting parents to take vacation days, with timing varying across German federal states due to different vacation schedules.

\subsubsection{Public holidays}

\begin{figure}[!htp]
    \centering    \includegraphics[width=0.5\textwidth]{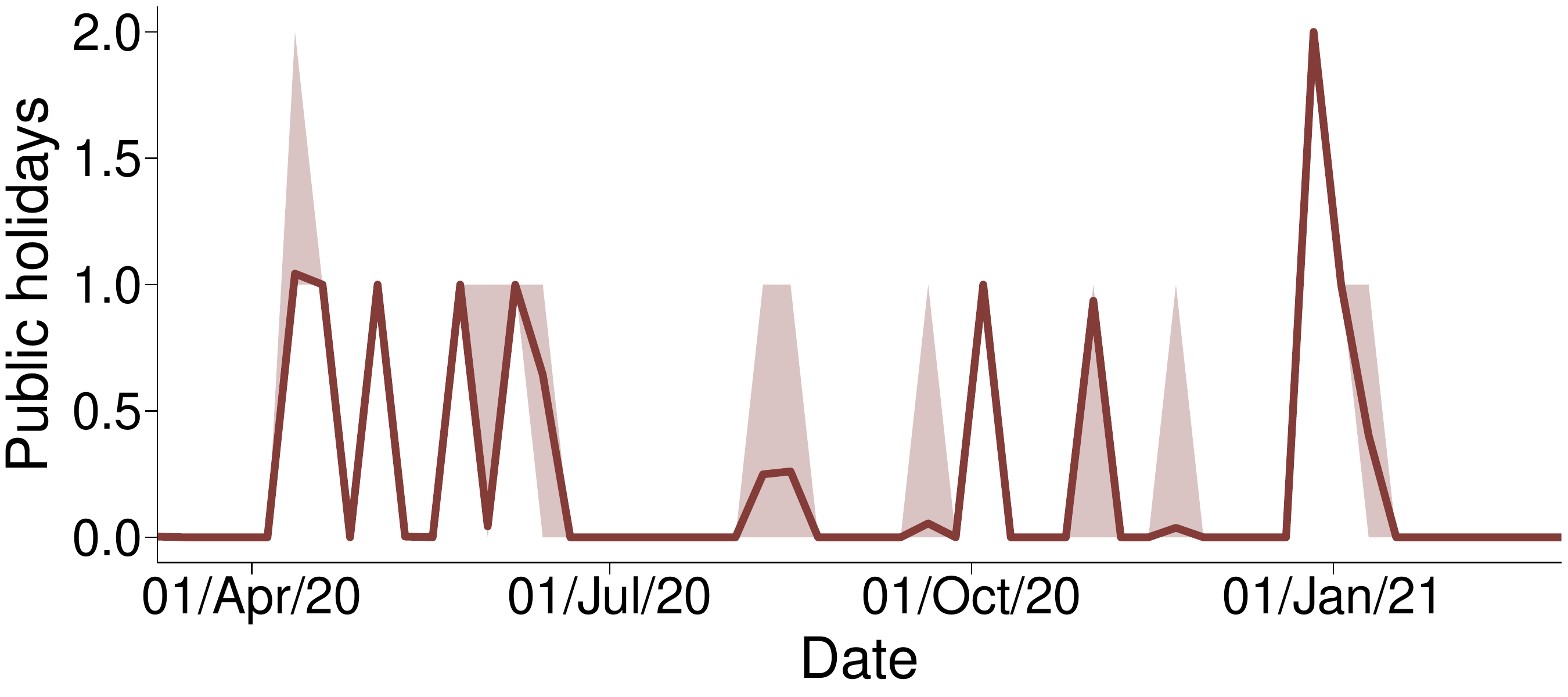}
    \caption{\textbf{With the exception of Christmas week, only one public holiday occurs per week.} For each week, the number of public holidays that fall on a weekday are summed up. Line represents the mean across federal states, ribbons represent the 95\% interpercentile range. The majority of public holidays occurs nation-wide.
    Only Christmas week has two public holidays; all other weeks contain a single holiday or none.}
    \label{fig:InputPubHol}
\end{figure}

Public holidays that fall on weekdays decrease the out-of-home duration both directly by exempting many employees from work and indirectly by encouraging many employees to take additional vacation days on the weekdays preceding and/or following the public holiday.
This decreases the out-of-home duration on the public holiday itself and the days adjacent to the public holiday, which, in turn, decreases the weekly average out-of-home duration.
In Germany, some public holidays are national holidays (including Easter~Monday, New~Year's~Eve, and New~Year's~Day), while others differ by federal state (including International~Women's~Day and All~Saints'~Day). 
For each week, we sum up the number of public holidays that fall on a weekday (Fig.~\ref{fig:InputPubHol}). 
During the study period, Christmas week was the only week with two public holidays.
Overall, public holidays decrease the weekly out-of-home duration both directly through work-free days and indirectly through adjacent vacation days, with effects varying by region due to differences in federal holiday schedules.

\FloatBarrier

\FloatBarrier
\subsubsection{COVID-19 Case Numbers}

\begin{figure}[!htp]
    \centering    \includegraphics[width=0.5\textwidth]{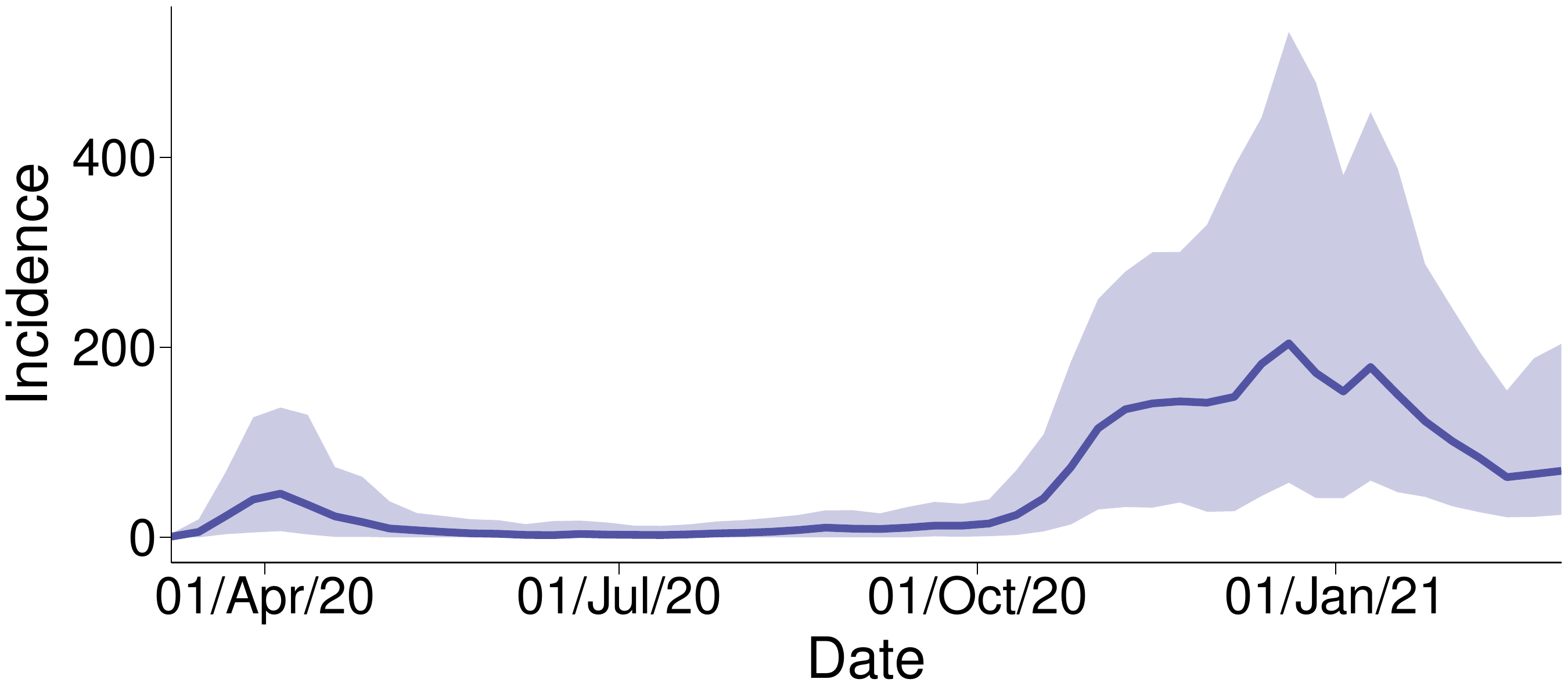}
    \caption{\textbf{The first COVID-19 wave was largely suppressed in Germany, higher incidences are reached during the second wave.} The line represents the mean of the 7-day incidence per 100.000 inhabitants across districts, ribbons represent the 95\% interpercentile range.
    Two local maxima occur: due to the first wave (spring 2020) and the second wave (winter 2020/2021).
    Peak heights differ widely across districts, with some districts experiencing almost no first wave.}
    \label{fig:InputCases}
\end{figure}

\begin{figure}[!htp]
    \centering    \includegraphics[width=0.80\textwidth]{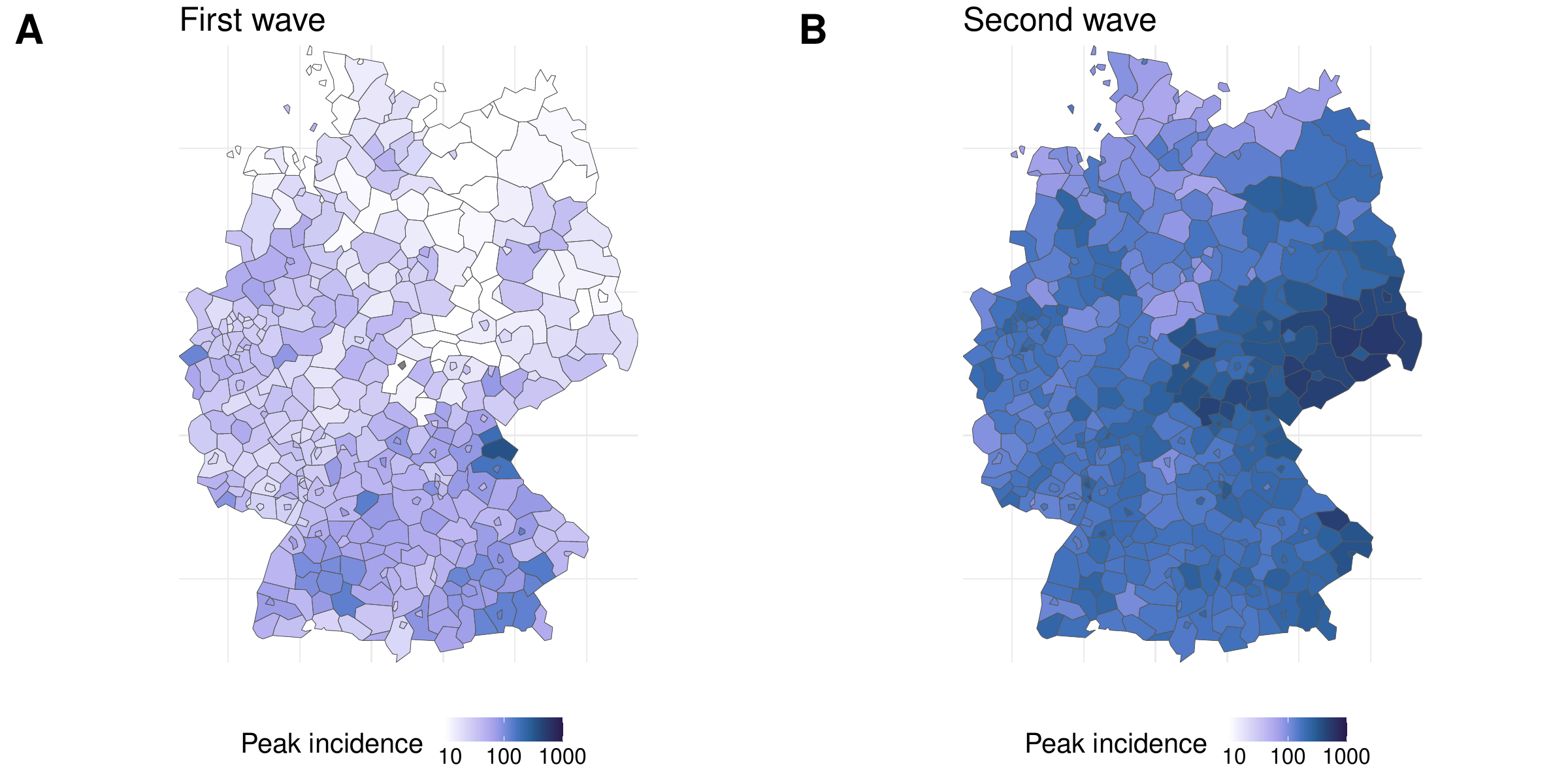}
    \caption{\textbf{Spatial depiction of COVID-19 incidences: southern/western Germany dominated the first wave, while eastern Germany (near Czechia) had the highest rates in the second wave.} For each district, the 90th percentile of the first and second wave are depicted.}\label{fig:CasesSpatial}
\end{figure}

In Germany, the first COVID-19 case was detected on January 27th, 2020. 
Case numbers began to steeply rise in March 2020, reaching the first local maximum between March~29th,~2020 and April~12th,~2020, depending on the district (Fig.~\ref{fig:InputCases}).
Across districts, the local maximum 7-day incidence/100,000 averaged 54 (median of maxima: 44, IQR: [27,67]).
The height of the initial wave differed greatly across districts, with some districts experiencing no first wave at all, and the highest outbreaks being observed in southern Germany and parts of North Rhine-Westphalia (Fig.~\ref{fig:CasesSpatial}~A).
From the maximum onwards, case numbers decreased and remained low throughout summer.
Cases began to rise again beginning mid-September, reaching a second peak around the turn of the year 2020/2021.
Again, second wave heights differed greatly across districts, with a mean maximum of 251 (median: 223, IQR: [181, 294]).
Eastern Germany, specifically the districts close to Czechia, endured the highest second waves (Fig.~\ref{fig:CasesSpatial}~B).
Overall, during the first pandemic year, Germany experienced two pandemic waves, one in the spring and one in the fall--winter, with large differences between districts.

\FloatBarrier
\subsection{Output Variable: Out-of-home Duration} \label{sec:SupplOutOfHome}

\begin{figure}[!htp]
    \centering    \includegraphics[width=0.5\textwidth]{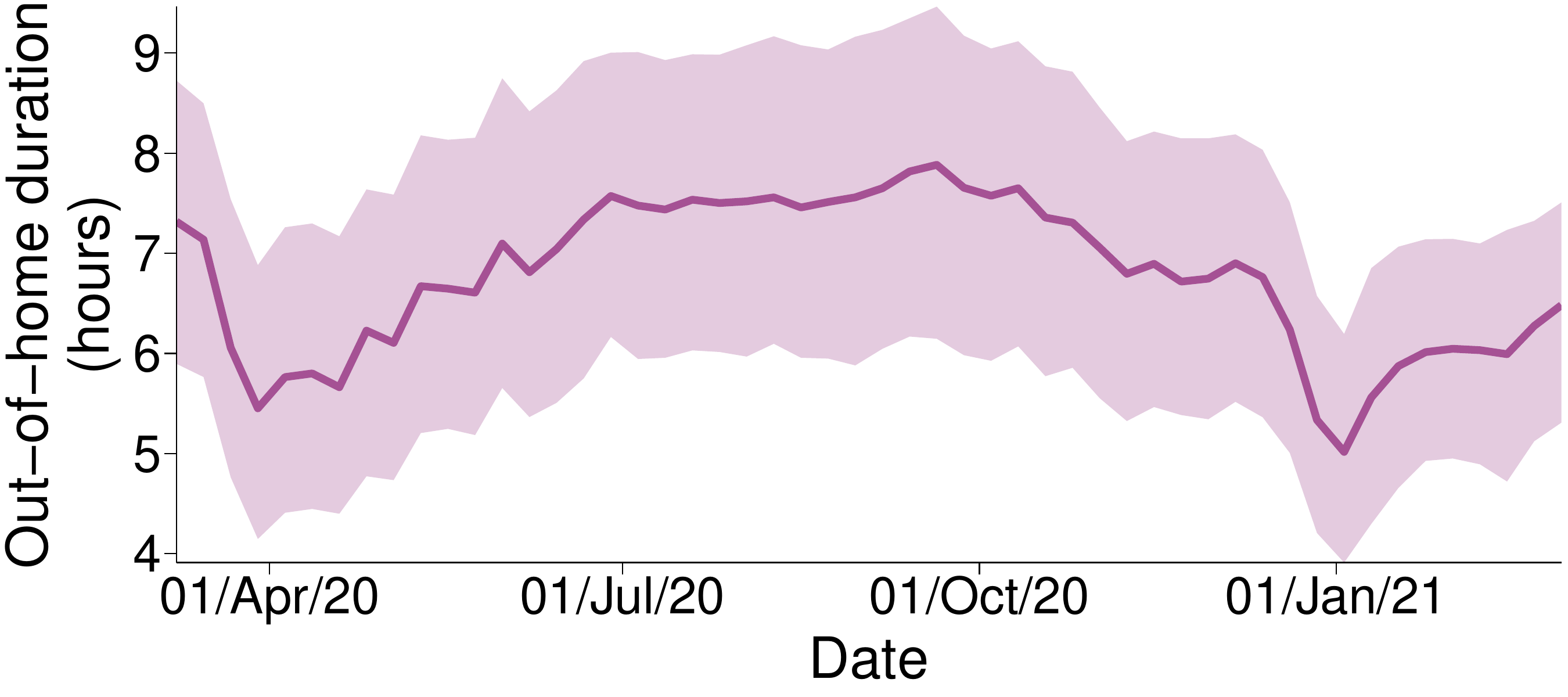}
    \caption{\textbf{In all districts, the out-of-home duration decreases sharply in spring and winter 2020/2021.} Line represents the mean across districts, ribbons represent the 95\% interpercentile range. Sharp decreases can be observed in March and December~2020. Out-of-home duration does not plateau at the local minima, but immediately rises again. During the summer 2020, out-of-home durations exceeding pre-pandemic levels from early March are observed.}\label{fig:InputOutOfHomeDuration}
\end{figure}

The out-of-home duration is a measure of local mobility, describing the amount of time (in hours and per day) an average person spends outside their home.
This duration fluctuated considerably throughout the study period, with the first change being observed in the second week of March~2020, when the out-of-home duration sharply decreased (Fig.~\ref{fig:InputOutOfHomeDuration}). 
Contact reductions were introduced in the week ending March~23rd,~2020, the week the out-of-home duration reached a local minimum. 
Rather than plateauing at this low point, the out-of-home duration began to increase immediately, notably before the first governmental relaxations (weeks ending May~4th,~2020 and May~11th,~2020). 
Throughout the summer of 2020, the out-of-home duration exceeded pre-pandemic levels from early March~2020, until mid-September, when another decrease in out-of-home duration is observed. 
The so-called ``lockdown light'', introduced on November~2nd,~2020, stopped the decreasing trend and lead to a plateau (potentially due to hasty Christmas-shopping and the looming stronger lockdown).
This plateau was followed by a dramatic decrease coinciding with more restrictive contact measures introduced on December~16th,~2020. 
During the winter of 2020/2021, the out-of-home duration fell to levels comparable to those observed during the initial COVID-19 wave in March~2020.
Altogether, the out-of-home fluctuated throughout the study period, declining sharply during lockdowns in March 2020 and winter 2020/2021, but exceeding pre-pandemic levels during summer 2020, with changes often occurring before rather than after policy implementations.

\FloatBarrier
\subsection{Bayesian hierarchical model}

\begin{figure}[!htp]
    \centering    \includegraphics[width=\textwidth]{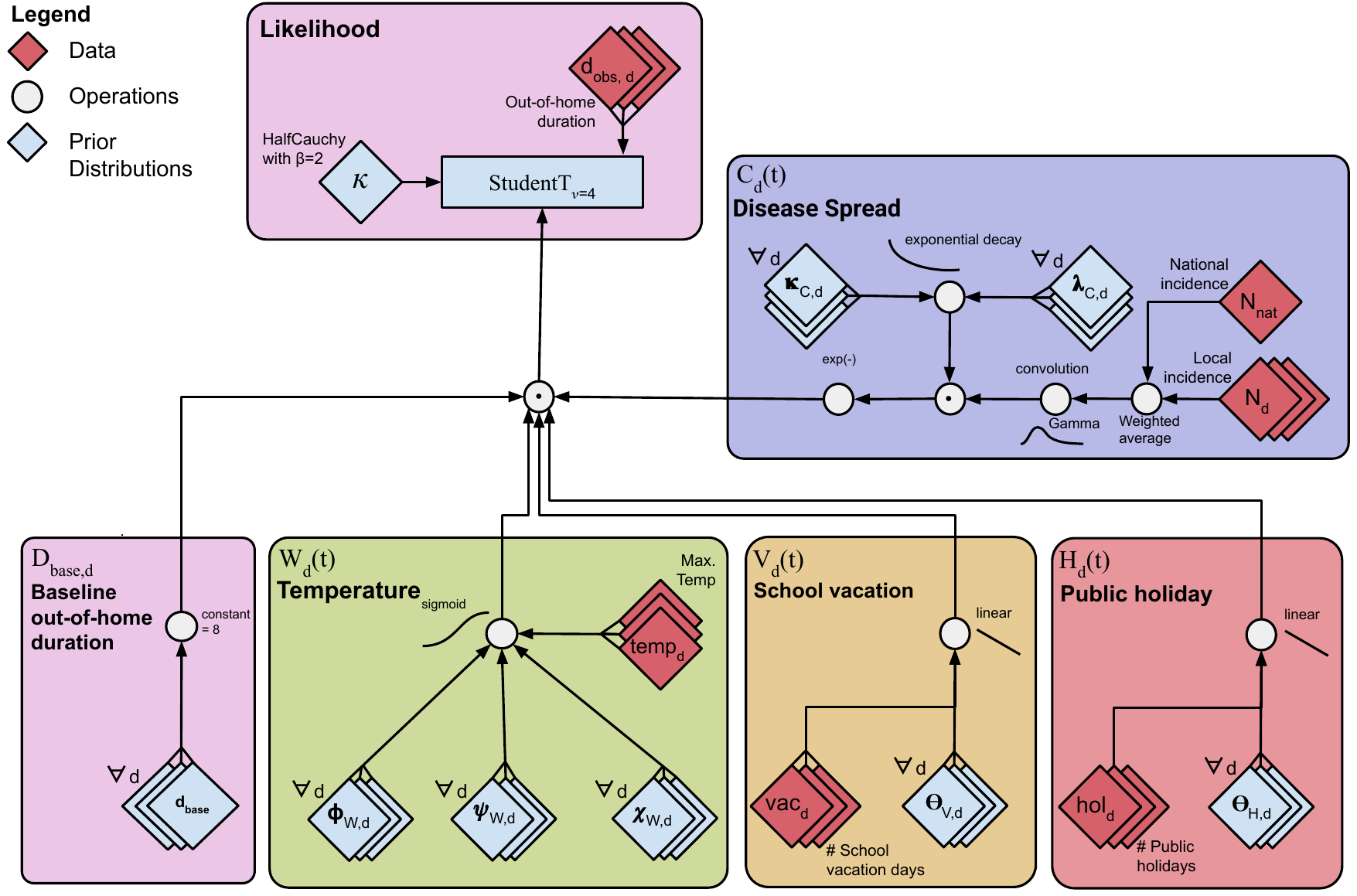}
    \caption{\textbf{Statistical model overview.}}
    \label{fig:model_overview}
\end{figure}

The model aims to infer the out-of-home duration $\out(t)$
for each German district $d \in \{1,\dots,400\}$ for each week $t \in \{1,\dots, 52\}$, spanning the period from March 2020 until March 2021.
We assume that the out-of-home duration $\out(t)$ can be modeled as the product of a baseline out-of-home duration $\dbase$, multiplied by a disease factor dependent on local and national COVID-19 spread $\disease(t)$ and 
 three disease-independent factors (temperature $\tempfac(t)$, school vacations $\vac(t)$, and public holidays $\hol$, Fig.~\ref{fig:model_overview} for graphical overview and Table~\ref{tab:Priors} for summary of prior distributions).
We assume that these factors are multiplicative (i.e. additive in log-space) as changes in one factor do not yield changes in another and as we assume that their effects can be expressed as multiplies of a base level. 
 The multiplicative inference model reads:

\begin{align} \label{Eq:oOH2020}
    \out(t) = \dbase \cdot \tempfac(t) \cdot \vac(t) \cdot \hol(t) \cdot \disease(t)
\end{align}

The different factors are discussed in the subsequent subsections.

\FloatBarrier

\subsubsection*{Baseline Out-of-home Duration}

The baseline out-of-home duration \dbase \ differs across districts:
Districts with high concentration of tradespeople, healthcare professionals, hospitality workers, or with a large share of employees working in manufacturing or mining, typically have longer out-of-home durations as employees need to by physically present onsite, forcing them to commute to work and leave their home.  
In contrast, districts dominated by remote work, with high shares of retirees or high unemployment rates, may show lower baseline out-of-home duration.
Finally, commuting time is typically longer in rural than in urban areas, as employees in rural areas need to cross greater distances to reach their worksite, increasing out-of-home duration \cite{Henning2018}. 
The baseline out-of-home duration \dbase \ encapsulates this district-dependency, representing the out-of-home duration during a week with no disease spread, no public holidays or school vacation, and average temperatures:
\begin{align*}
\dbase &= \theta_{\text{base},d} \cdot 8, \\
\text{where } \theta_{\text{base},d} &= \text{exp}(\mu^{\theta}_{\text{base}} + z_{\text{base},d}\cdot \sigma^{\theta}_{\text{base}}), \\
z_{\text{base},d} &\sim \mathcal{N}(0, 1), \\ 
\mu_{\text{base}}^\theta &\sim \mathcal{N}(0, 0.2), \\
\sigma_{\text{base}}^\theta &\sim \text{HalfNormal}(0.3).
\end{align*}
The underlying assumption of \dbase \ is that the baseline out-of-home duration fluctuates around 8 hours, reflecting the typical daily work time of a full-time employee in Germany.

\subsubsection*{Temperature}

Weather --- here represented by temperature --- increases out-of-home duration during the warmer months (typically April through September) and decreases out-of-home duration during the colder months (typically October through March).
We model this effect using a sigmoid function, which captures the saturation of weather's effect at the temperature extremes.
The inflection point of the sigmoid takes on a value of 1, representing a week with average annual temperature, where temperature neither in- nor decreases out-of-home duration.
\begin{align*}
  \tempfac (t) &= \amplitudetemp \cdot\left(\frac{1}{1 + exp(- (\temp(t) - \shifttemp)/\slopetemp)}\right) + \bigg(1 - \amplitudetemp/2\bigg), \\
\amplitudetemp &= \mu_W^\phi + \sigma_{W}^\phi \cdot z^{\phi}_{W,d}, \\
z_{W,d}^{\phi}&\sim \mathcal{N}(0,1), \\
\mu_W^\phi &\sim \text{HalfCauchy}(1), \\
\sigma_W^\phi &\sim \text{Exp}(10), \\
\shifttemp &= \mu_W^\psi + \sigma_W^\psi \cdot  z^{\psi}_{W,d}, \\
z^{\psi}_{W,d} &\sim \mathcal{N}(0,1 ), \\
\mu_W^\psi &\sim \mathcal{N} (15,3), \\
\sigma_W^\psi &\sim \text{Exp} (10), \\
\slopetemp &= \text{exp}(\mu_W^\chi + \sigma_W^\chi \cdot z^{\chi}_{W,d}), \\
z^{\chi}_{W,d}&\sim \mathcal{N}(0,1), \\
\mu_W^\chi &\sim \mathcal{N}(\text{ln}(4),0.5), \\
\sigma_W^\chi &\sim \text{Exp}(10). \\
\end{align*}

\subsubsection*{School Vacations}
School vacations typically decrease out-of-home duration because students no longer attend schools and parents often take vacation days or work remotely to care for their children.
The more business days are school vacation days, the stronger the reduction in week $t$.
\begin{align*}
\vac(t) &= \frac{\thetavac-1}{7} \cdot \left(\#\text{School vacation days during week }t\right) + 1, \\
\text{where } \thetavac &= \mu^\theta_V + \sigma^\theta_V \cdot z_{V,d}, \\
z_{V,d} &\sim \mathcal{N}(0,1), \\
\mu^\theta_V &\sim \mathcal{U}[0.8, 1.0], \\
\sigma_V^\theta &\sim \text{Exp}(10).
\end{align*}
In Germany, students go to school from Monday to Friday. 
Consequently, ``\#School vacation days during week t'' may take up integer values between 0 and 5. 
If every school day of week $t$ is a vacation day, then $\vac(t) = \frac{5}{7}\thetavac + \frac{2}{7}$, reflecting that only the out-of-home duration on the five potential school days is affected by the school vacation, while the out-of-home duration on the weekend remains unaffected.

\subsubsection*{Public Holidays}
Public holidays that fall on weekdays decrease out-of-home duration both by exempting employees from work and by incentivizing them to extend their time of by taking additional vacation days before and/or after the holiday.

\begin{align*}
\hol(t) &= \frac{\thetahol-1}{7} \cdot (\#\text{Public holidays during week }t) + 1, \\
\text{where } 
\thetahol &= \mu_H^\theta + \sigma_H^\theta \cdot z_{H,d}, \\
z_{H,d} &\sim \mathcal{N}(0,1), \\
\mu_H^\theta &\sim \mathcal{U} [0.9, 1], \\
\sigma_H^\theta &\sim \text{HalfNormal} (0.25).
\end{align*}
In Germany, the majority of full time employees works from Monday to Friday. 
Consequently, ``\#Public holidays during week $t$'' may take on integer values between 0 and 5. 
If every weekday of week $t$ is a public holiday, then $\hol(t) = \frac{5}{7}\thetahol + \frac{2}{7}$, reflecting that only the out-of-home duration on the weekdays is affected by the publich holiday, while the out-of-home duration on the weekend remains unaffected.

\subsubsection*{Disease Spread} \label{sec:MethodsDisease}

We assume that the effect of disease spread on out-of-home duration to have a decreasing impact: High case numbers motivate more stringent NPIs and increased risk perception which both, in turn, lead to smaller out-of-home activity and consequently decrease out-of-home duration.
The disease factor $\disease(t)$ takes the national and local COVID-19 7-Day incidence rate as input and is computed in four steps:
\begin{enumerate}
\item First, assuming that both local and national disease spread influence local mobility behavior, a weighted average of the normalized local COVID-19 incidence rate $N_d(t)$ and normalized national COVID-19 $N_{\text{nat}}(t)$ incidence rate is computed.
\item Second, assuming that out-of-home duration in week $t$ is not only instantaneously influenced by disease spread in week $t$, but also by disease spread in the recent past, the weighted average is convolved with a Gamma distribution.
\item Third, assuming that the population's perception of risk decreases over time as pandemic fatigue sets in, the convolved case numbers are multiplied with an exponential decay. This ensures that the same level of case case numbers reduces the out-of-home duration less at a later point in time.
\item Fourth, as we assume it has a multiplicative effect, we take the exponential of the convoluted and exponentially decayed case numbers. This also ensures a smaller than 1 multiplicative effect for positive case numbers.

\end{enumerate}
Together the disease factor reads:
\begin{align}
\disease(t) &= exp\bigg(-z_{C,d}(t) \cdot \int_{-\infty}^t ( \omega_{C,d}  \cdot N_{d}(t) + (1-\omega_{C,d} ) \cdot N_{\text{nat}}(t)) \cdot \mathcal{G}(t-t', \mu_{C,d}, \sigma_{C,d})dt'\bigg),
\end{align}
where \amplitudedisease is the exponential decay:
\begin{align}
z_{C,d}(t) &= \amplitudedisease \cdot exp(-t/\slopedisease),\\
\amplitudedisease &= \text{exp}(\mu_C^{\phi} + \sigma_C^{\phi} \cdot z^{\kappa}_{C,d}), \\
z^{\kappa}_{C,d}&\sim \mathcal{N}(0,1) ,\\
\mu_C^{\phi} &\sim \mathcal{N}(\text{ln}(1.5), 0.25), \\
\sigma_C^{\phi} &\sim \text{Exp}(10), \\
\slopedisease &= \text{softplus}(z^{\lambda}_{C,d}), \\
z^{\lambda}_{C,d} &\sim \mathcal{N}(\mu_C^{\psi}, \sigma_C^{\psi}), \\
\mu_{C}^{\psi} &\sim \mathcal{N}(30, 1), \\
\sigma_C^{\psi} &\sim \text{HalfNormal}(0.1).
\end{align}
$\omega_{C,d}$ is the local vs national weighting:
\begin{align}
\omega_{C,d} &= 1/(1+exp(- z^{\omega}_{C,d})), \\
z^{\omega}_{C,d} &\sim \mathcal{N} (\mu_{C}^{\omega}, 1), \\
\mu_C^{\omega} &\sim \mathcal{N}(0, 0.5).
\end{align}
And $\mu_{C,d}$ and $\sigma_{C,d}$ the mean and standard deviation of the Gamma kernel:
\begin{align}
\mu_{C,d} &= \text{softplus}(\mu_{C}^{\mathcal{G}} + \sigma_{C}^{\mathcal{G}} \cdot z^{\Gamma}_{C,d} ), \\
z^{\Gamma}_{C,d} &\sim \mathcal{N}(0,1), \\
\mu_{C}^{\mathcal{G}} &\sim \mathcal{N}(3, 1), \\
\sigma_{C}^{\mathcal{G}} &\sim \text{HalfNormal}(0.25), \\
\sigma_{C,d} &= \mu_{C,d} / \sqrt{\alpha^{\mathcal{G}}_{C}},\\
\alpha_{C}^{\mathcal{G}} & \sim \text{LogNormal}(\text{ln}(3), 0.125).
\end{align}

\subsubsection*{Likelihood}

Finally, we define a goodness of fit of our model to the observed out-of-home duration. 
The likelihood is hereby modeled using a Student's $t$-distribution as this allows for some outliers due to its heavier tails.
The likelihood reads
\begin{align*}
L_d(t) &\sim \text{StudentT}_{\nu=4}(\out(t), \sigma_L), \\
\sigma_L &\sim \text{HalfCauchy}(2).
\end{align*}

    \begin{xltabular}{\textwidth}{|l|l|X|} 
    \hline
    \textbf{Parameter}   & \textbf{Prior Distribution} & \textbf{Rationale/Meaning} \\ \hline \hline
    \multicolumn{3}{|l|}{Baseline out-of-home duration \dbase} \\ \hline
    $\mu^{\theta}_{\text{base}}$ & $\mathcal{N}(0,0.2)$ & Used to determine how much \dbase \ in district $d$ deviates from the assumed 8 hours of baseline out-of-home duration\\
    $\sigma^{\theta}_{\text{base}}$  & HalfNormal(0.3) & Used to determine variance across districts in deviating from the assumed 8 hours of baseline out-of-home duration, weakly-informative \\ \hline \hline
    \multicolumn{3}{|l|}{Temperature factor \tempfac} \\ \hline 
    $\mu^{\phi}_{W}$ & HalfCauchy(1) & Determines if temperature impact out-of-home duration. For values close to 0, the impact of temperature vanishes \\
    $\sigma^{\phi}_{W}$ & Exp(10) & Determines variance across districts in impacting out-of-home duration, weakly-informative\\
    $\mu^{\psi}_{W}$ & $\mathcal{N}(15,3)$ & Determines shift of sigmoid function. Based on previous work, in which we explored the threshold temperature above which individuals move their activities outside \cite{Mueller2021, Paltra2024} \\
    $\sigma^{\psi}_{W}$ & Exp(10) & Determines variance in shift across districts, weakly-informative \\
    $\mu^{\chi}_{W}$ & $\mathcal{N}(\text{ln}(4),0.5)$ & Determines slope of sigmoid function, weakly-informative \\
    $\sigma^{\chi}_{W}$ &  Exp(10) & Determines variance in slope across districts, weakly-informative \\ \hline \hline
    \multicolumn{3}{|l|}{School vacation factor \vac} \\ \hline
    $\mu^{\theta}_{V}$  & $\mathcal{N}[0.8,1]$ & Amplitude of impact of vacation factor, non-informative \\
    $\sigma^{\theta}_{V}$  & Exp(10) & Determines variance in amplitude across districts, weakly-informative \\ \hline \hline
    \multicolumn{3}{|l|}{Public holiday factor \hol} \\ \hline
    $\mu^{\theta}_{H}$  & $\mathcal{U}[0.9,1]$ & Amplitude of impact of public holiday factor, non-informative \\
    $\sigma^{\theta}_{H}$ & HalfNormal(0.25) & Determines variance in amplitude across districts \\ \hline \hline
    \multicolumn{3}{|l|}{Disease factor \disease} \\ \hline
    $\mu^{\phi}_C$ & $\mathcal{N}(\text{ln}(1.5),0.25)$ & Determines initial quantity of exponential decay, weakly-informative \\
    $\sigma^{\phi}_C$ & Exp(10) & Determines variance in initial quantity across districts, weakly-informative \\
    $\mu^{\psi}_C$ & $\mathcal{N}(30,1)$ & Determines exponential decay constant, weakly-informative \\
    $\sigma^{\psi}_C$ & HalfNormal(0.1) & Determines variance in exponential decay constant across districts \\
    $z_{C,d}^{\omega}$ & $\mathcal{N}(\mu_C^{\omega},1)$ & Determination of weight of normalized local incidence\\
    $\mu_C^{\omega}$ & $\mathcal{N}(0,0.5)$ & Determines mean of weight of normalized local incidence\\
    $\mu_{C}^{\mathcal{G}}$ & $\mathcal{N}(3,1)$ & Determines mean of gamma memory distribution\\
    $\sigma_{C}^{\mathcal{G}}$ & HalfNormal(0.25) & Determines variance in mean across districts \\
    $\alpha_C^{\mathcal{G}}$ & LogNormal(ln(3),0.125) & Used (in combination with $\mu_{C,d}$) to determine standard deviation of gamma memory distribution. Based on the assumption that districts differ little in their memory. \\ \hline
    \caption{\textbf{Prior distributions and their meaning in our Bayesian inference model.}}
        \label{tab:Priors}
    \end{xltabular}

\FloatBarrier

\subsection{Inference}
\label{sec:methods:inference}

We use the package PyMC \cite{Abril2023} to build the model and the NUTS sampler for inference \cite{Hoffman2014} via the library Nutpie \cite{Seyboldt2024}. We use 2000 tuning steps and 1000 draws in 4 parallel chains. We made sure the chains converged, as the R-hat value is below 1.07. 

\subsection{Multiple Regression}

We explore differences across districts for three outcome variables: reaction strength, weight of local incidence, and pandemic fatigue. 
To this end, we build a multiple linear regression model for each outcome variable, considering population density as well as eighteen demographic, socio-economic, and political factors as explanatory variables (Table~\ref{tab:RegressionVariableOverview} for overview of considered variables).

\begin{longtable}{|p{3.2cm}|p{6cm}lll|}
\hline
        \textbf{Variable} & \textbf{Description} & \textbf{Mean} & \textbf{Median} & \textbf{IQR} \\ \hline
        \raggedright Population density & Extrapolation of inhabitants per km$^2$ on December~31st,~2022 based on 2011 census  & 544.1 & 204.5 & [118.0, 686.0] \\
        Income & Primary income of private households, including private non-profit organizations in 2020, in EURO & 27811.9 & 27842.5 & [24679.0, 31021.3] \\
        Average age & Average age of the population on December~31st,~2020 & 45.2 & 45.0 & [44.0, 46.4] \\
        \raggedright 65+ year olds & Share of people aged 65 and older in the total population in 2020, in \% & 22.3 & 22.0 & [20.5, 23.9] \\
         \raggedright Small children in childcare & Share of children under the age of 3 in daycare facilities/daycare centers in the age group in 2020, in \% & 33.1 & 29.9 & [24.5, 36.3] \\                   
        Unemployment rate & Unemployment rate relative to the total civilian labor force (sum of working population and registered unemployed persons) in 2020, annual average, in \% & 5.5 & 5.2 & [3.8, 6.6] \\
        Employment rate & Division of the number of employees subject to social insurance contributions on June~30th,~2020 by the population aged 15-64 on December 31 of the previous year, in \%. The 2020 \emph{employment rate} was unavailable for the following districts: Suhl, Wartburgkreis, Schmalkalden-Meiningen, Ilm-Kreis, Sonneberg, Saalfeld-Rudolstadt. For these districts, we used the 2021 \emph{employment rate} & 62.2 & 62.7 & [60.0, 65.3] \\
         Service sectors & Proportion of employed persons in service sectors in 2020, in \% & 70.7 & 70.4 & [63.7, 77.7] \\
                  Manufacturing sector & Proportion of employed persons in the manufacturing sector in 2020, in \% & 27.4 & 26.4 & [20.3, 34.1] \\
         TTHIC sectors & Proportion of employed persons in the trade, transport, hospitality, information, and communication sector in 2020, in \% & 24.4 & 24.0 & [21.8, 26.9] \\  
         Finance sector & Proportion of employed persons in financial, insurance, and business services, real estate and housing sector in 2020, in \% & 14.1 & 13.1 & [11.2, 15.7] \\     
         Construction & Proportion of employed persons working in construction in 2020, in \% & 6.7 & 6.6 & [5.0, 8.3] \\
    \raggedright Agriculture, forestry, fisheries & Proportion of employed persons in agriculture, forestry and fisheries in 2020, in \% & 1.9 & 1.6 & [0.5,2.9] \\
    Voter turnout & Federal election turnout in 2021 in districts, in \% & 76.3 & 76.5 & [74.0, 79.1] \\
    CDU & Share of votes for political party \emph{Christian Democratic Union of Germany} (CDU) in 2021 federal election, in \% & 25.0 & 24.5 & [20.4, 29.7] \\
    SPD & Share of votes for political party \emph{Social Democratic Party of Germany} (SPD) in 2021 federal election, in \% & 25.7 & 25.7 & [21.0, 29.8] \\
    AfD & Share of votes for political \emph{Alternative for Germany} (AfD) in 2021 federal election, in \% & 11.4 & 9.5 & [7.7, 12.6] \\
    FDP & Share of votes for political party \emph{Free Democratic Party} (FDP) in 2021 federal election, in \% & 11.1 & 10.9 & [9.5, 12.2] \\
    Green party & Share of votes for political party \emph{Alliance 90/The Greens} (green party) in 2021 federal election, in \% & 12.8 & 12.5 & [9.1, 15.9] \\
    \hline
\caption{\textbf{Demographic, socioeconomic, and political variables considered in the regression models differ across districts.} Due to the differences in scale, all variables are standardized for the regressions.}
\label{tab:RegressionVariableOverview}
\end{longtable}

\FloatBarrier

\FloatBarrier

\subsection{Linear Model Explaining Cross-District Variation in Reaction Strength} \label{sec:RegressionReactionStrength}

First analysis reveals significant differences in reaction strength between district types (Section~\ref{sec:ReacStrengthDistType}).
However, district type -- i.e. population density -- alone does not suffice to fully explain the observed differences.
To explore the differences in reaction strength further, we conduct a multi-step analysis separated by wave.
To this end, we define \emph{reaction strength first wave} as the integral over the exponential decay function for the first 13 weeks of the study period, ranging from the week ending March~8th,~2020 until the week ending May~31st,~2020, while we define \emph{reaction strength second wave} as the integral over the exponential decay function for the final 26 weeks of the study period, ranging from the week ending September~6th,~2020 until the week ending February~28th,~2021.
For each wave we follow the same four step approach:
\begin{enumerate}
    \item Exploratory analysis of correlation coefficients of \emph{reaction strength} and \emph{population density} and eighteen additional demographic, socioeconomic, and political explanatory variables (Table~\ref{tab:RegressionVariableOverview} for considered variables).
    \item Exhaustive search for each number of variables between 1 and 19. For each number of variables, we chose the subset with the highest adjusted $R^2$.
    \item Model selection across subset sizes using (adjusted and predicted) $R^2$, Mallow's $C_p$, AIC, and BIC.
    \item Regression on \emph{reaction strength} to determine the demographic, socioeconomic, and political variables' ability to explain observed variance in \emph{reaction strength}.
\end{enumerate}

\subsubsection{Exploratory Analysis}

\FloatBarrier
\begin{figure}[!htp]
    \centering       
    \includegraphics[width=0.47\textwidth]{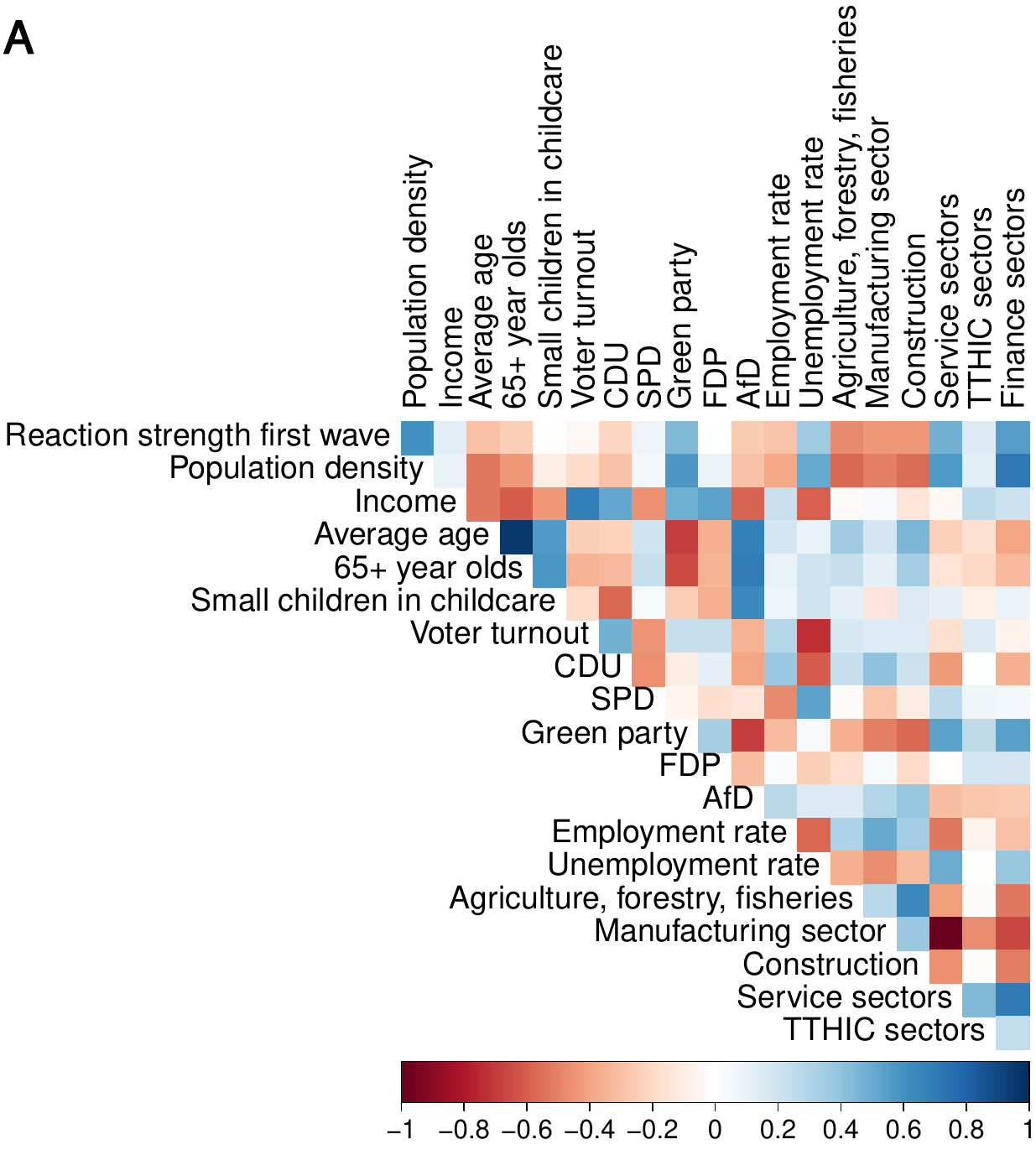}
    \includegraphics[width=0.49\textwidth]{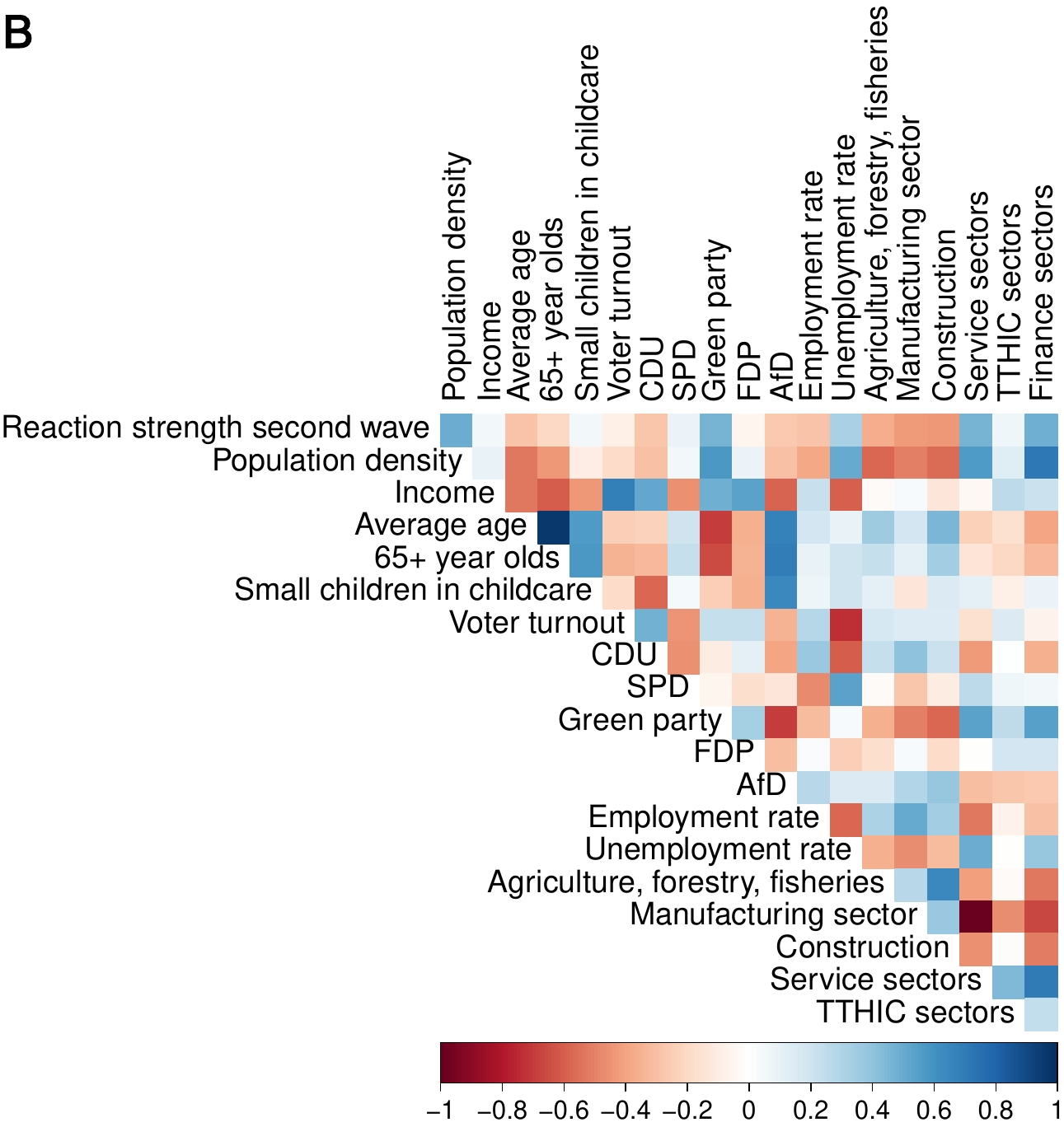}
    \caption{\textbf{\emph{Reaction strength} of either wave is strongly correlated with \emph{population density} and employment sectors which easily or not at all permit remote work.} \emph{Reaction strength} correlates most strongly positively with \emph{population density}, \emph{service sectors}, and \emph{finance sector}. \emph{Agriculture, forestry, fisheries}, \emph{manufacturing sector}, and \emph{construction} are negatively correlated with \emph{reaction strength}.
    \textbf{A.} Correlation coefficients of \emph{reaction strength first wave} and the 18 chosen explanatory variables.
    \textbf{B.} Correlation coefficients of \emph{reaction strength second wave} and the 18 chosen explanatory variables. Only first the first row differs from panel A.}
    \label{fig:Regression_Corr}
\end{figure}

In the first analysis step, we consider linear correlations between \emph{reaction strength first wave} and \emph{reaction strength second wave} on the one hand, and the explanatory variables on the other hand.
\emph{Population density} shows the largest positive correlation with \emph{reaction strength first wave} (0.60).
The \emph{finance sector} (0.55) and the \emph{service sectors} (0.48) exhibit the second and third largest positive correlations, respectively.
A possible explanation may be that the finance sector is dominated by white collar jobs which allow for remote work more easily.
In contrast, the service sectors were completely shut down for large parts of the COVID-19 pandemic, forcing employees to stay home and consequently explaining the strong positive correlation.
Finally, \emph{green party} is also strongly positively correlated with \emph{reaction strength first wave} (0.44).
In contrast, employment sectors that do not allow any remote work and necessitate on-site labor, are strongly negatively correlated with \emph{reaction strength first wave}: \emph{agriculture, forestry, fisheries} (correlation coefficient of $-0.47$), \emph{construction} ($-0.44$), and the \emph{manufacturing sector} ($-0.43$).
Examining the correlation coefficients among the explanatory variables reveals an extremely positive correlation between \emph{average age} and \emph{65+ year olds} (0.97).
\emph{Manufacturing sector} and \emph{service sectors} are strongly negatively correlated ($-0.99$), and so are \emph{voter turnout} and \emph{unemployment rate} ($-0.74$).
We further observe a strong negative correlation between \emph{green party} and the right-wing party \emph{AfD} ($-0.70$).

Examining \emph{reaction strength second wave} yields similar patterns, though with slightly smaller correlation coefficients.
Again, \emph{reaction strength second wave} is strongly positively correlated with \emph{population density} (0.50), the \emph{finance sector} (0.48), \emph{green party} (0.47), and the \emph{service sectors} (0.46).
Again, we observe strong negative correlations with \emph{construction} ($-0.44$), the \emph{manufacturing sector} ($-0.43)$, \emph{agriculture, forestry, fisheries} ($-0.36$).

Overall, the correlation analysis reveals that reaction strength in both waves is consistently positively correlated to population density, sectors amenable to remote work or disrupted by lockdowns, and green party support, while being inversely related to industries requiring on-site labor.

\FloatBarrier

\subsubsection{Exhaustive Search} \label{sec:RecStrengthBestSubset}
We perform an exhaustive search of all possible variable combinations to determine which independent variables predict \emph{reaction strength first wave} and \emph{reaction strength second wave}. 
For this analysis step, we use the R package olsrr \cite{Hebbali2024}. 
For each subset size (1--19 variables), we select the model that maximized adjusted $R^2$ (for first wave analysis, see Table~\ref{tab:ExhSearchVariablesFirstWave} for variables chosen for each variable count and for second wave analysis, see Table~\ref{tab:ExhSearchVariablesSecondWave}).
For the first wave, via this method, \emph{population density}, the explanatory variable most strongly correlated with \emph{reaction strength first wave}, is chosen for every subset size, while \emph{unemployment rate}, the explanatory variable second-most-strongly correlated with \emph{reaction strength first wave}, is chosen from a subset size of 4 onward.
Overall, apart from population density, the variables chosen in the smaller models, mostly relate to different job sector and the necessity (or lack thereof) of on-site work: 
A high \emph{unemployment rate}
Working in the \emph{finance sector} as well as having a higher \emph{income} correlates with pursuing a white collar job, allowing for home office and remote work more readily.
Contrarily, working in \emph{agriculture, forestry, and fisheries} forces employees to work on-site.
For the second wave, via this method, political variables gained importance, with at least one political party variable select from the two variable model onward. 
Conversely, \emph{population density} is only chosen in the one variable model, the ten variables model, and from the twelve variables model onward.

\begin{longtable}{|p{1.7cm}|p{14cm}|}
\hline
\raggedright \textbf{No. of Variables} & \textbf{Variables chosen} \\ \hline
1 & Population density \\
2 & Population density, Finance sector \\
3 & Population density, Agriculture, forestry, fisheries, Manufacturing sector \\
4 & Population density, Voter turnout, Unemployment rate, Agriculture, forestry, fisheries \\
5 & Population density, Voter turnout, Unemployment rate, Agriculture, forestry, fisheries, Finance sector \\
6 & Population density, Voter turnout, Income, FDP, Unemployment rate, Agriculture, forestry, fisheries \\
7 & Population density, Voter turnout, Income, FDP, Unemployment rate, Agriculture, forestry, fisheries, Finance sector \\
8 & Population density, Voter turnout, Income, Green party, FDP, Average age, Unemployment rate, Agriculture, forestry, fisheries \\
9 & Population density, Voter turnout, Income, Green party, FDP, Average age, Unemployment rate, Agriculture, forestry, fisheries, Finance sector \\
10 & Population density, Voter turnout, Income, Green party, FDP, Average age, Unemployment rate, Agriculture, forestry, fisheries, TTHIC sectors, Finance sector \\
11 & Population density, Voter turnout, Income, Green party, FDP, 65+ year olds, Average age, Unemployment rate, Agriculture, forestry, fisheries, TTHIC sectors, Finance sector \\
12 & Population density, Voter turnout, Income, Green party, FDP, 65+ year olds, Average age, Unemployment rate, Agriculture, forestry, fisheries, Construction, TTHIC sectors, Finance sector \\
13 & Population density, Voter turnout, Income, Green party, FDP, 65+ year olds, Average age, Employment rate, Unemployment rate, Agriculture, forestry, fisheries, Construction, TTHIC sectors, Finance sector \\
14 & Population density, Voter turnout, Income, CDU, Green party, FDP, 65+ year olds, Average age, Employment rate, Unemployment rate, Agriculture, forestry, fisheries, Construction, TTHIC sectors, Finance sector \\
15 & Population density, Voter turnout, Income, Small children in childcare, CDU, Green party, FDP, 65+ year olds, Average age, Employment rate, Unemployment rate, Agriculture, forestry, fisheries, Construction, TTHIC sectors, Finance sector \\
16 & Population density, Voter turnout, Income, Small children in childcare, CDU, SPD, Green party, FDP, 65+ year olds, Average age, Employment rate, Unemployment rate, Agriculture, forestry, fisheries, Construction, TTHIC sectors, Finance sector \\
17 & Population density, Voter turnout, Income, Small children in childcare, CDU, SPD, Green party, FDP, AfD, 65+ year olds, Average age, Employment rate, Unemployment rate, Agriculture, forestry, fisheries, Construction, TTHIC sectors, Finance sector \\
18 & Population density, Voter turnout, Income, Small children in childcare, CDU, SPD, Green party, FDP, AfD, 65+ year olds, Average age, Employment rate, Unemployment rate, Manufacturing sector, Service sectors, Construction, TTHIC sectors, Finance sector \\
19 & Population density, Voter turnout, Income, Small children in childcare, CDU, SPD, Green party, FDP, AfD, 65+ year olds, Average age, Employment rate, Unemployment rate, Agriculture, forestry, fisheries, Manufacturing sector, Service sectors, Construction, TTHIC sectors, Finance sector \\ \hline
\caption{\textbf{Regression on reaction strength of \emph{first wave} favors population strength and demographic variables.} For each subset size (1--19 variables), all possible combinations are considered via an exhaustive search and the subset with the maximal adjusted $R^2$ is selected.}
\label{tab:ExhSearchVariablesFirstWave}
\end{longtable}

\begin{longtable}{|p{1.7cm}|p{14cm}|}
\hline
\raggedright \textbf{No. of Variables} & \textbf{Variables chosen} \\ \hline
1 & Population density \\
2 & Green party, Unemployment rate \\
3 & Green party, FDP, Finance sector \\
4 & Green party, FDP, Unemployment rate, Finance sector \\
5 & Green party, FDP, Unemployment rate, Construction, Finance sector \\
6 & CDU, SPD, FDP, AfD, Unemployment Rate, Finance sector \\
7 & CDU, SPD, FDP, AfD, Unemployment Rate, Construction, Finance sector \\
8 & CDU, SPD, FDP, AfD, 65+ year olds, Average age, Unemployment Rate, Finance sector \\
9 & Income, Small children in childcare, CDU, SPD, FDP, AfD, 65+ year olds, Average age, Unemployment rate \\
10 & Population density, Income, Small children in childcare, CDU, SPD, FDP, AfD, 65+ year olds, Average age, Unemployment rate \\
11 & Income, Small children in childcare, CDU, SPD, FDP, AfD, 65+ year olds, Average age, Employment rate, Unemployment rate, Agriculture, forestry, fisheries \\
12 & Population density, Income, Small children in childcare, CDU, SPD, FDP, AfD, 65+ year olds, Average age, Employment rate, Unemployment rate, Agriculture, forestry, fisheries \\
13 & Population density, Income, Small children in childcare, CDU, SPD, FDP, AfD, 65+ year olds, Average age, Employment rate, Unemployment rate, Agriculture, forestry, fisheries, TTHIC sectors \\
14 & Population density, Income, Small children in childcare, CDU, SPD, FDP, AfD, 65+ year olds, Average age, Employment rate, Unemployment rate, Agriculture, forestry, fisheries, TTHIC sectors, Finance sectors \\
15 & Population density, Income, Small children in childcare, CDU, SPD, FDP, AfD, 65+ year olds, Average age, Employment rate, Unemployment rate, Agriculture, forestry, fisheries, Construction, TTHIC sectors, Finance sectors \\
15 & Population density, Income, Small children in childcare, CDU, SPD, FDP, AfD, 65+ year olds, Average age, Employment rate, Unemployment rate, Agriculture, forestry, fisheries, Manufacturing sector, Construction, TTHIC sectors, Finance sectors \\
16 & Population density, Income, Small children in childcare, CDU, SPD, FDP, AfD, 65+ year olds, Average age, Employment rate, Unemployment rate, Manufacturing sector, Construction, Service sectors, TTHIC sectors, Finance sectors \\
17 & Population density, Income, Small children in childcare, CDU, SPD, FDP, AfD, 65+ year olds, Average age, Employment rate, Unemployment rate, Agriculture, forestry, fisheries, Manufacturing sector, Construction, Service sectors, TTHIC sectors, Finance sectors \\
18 & Population density, Voter turnout, Income, Small children in childcare, CDU, SPD, FDP, AfD, 65+ year olds, Average age, Employment rate, Unemployment rate, Agriculture, forestry, fisheries, Manufacturing sector, Construction, Service sectors, TTHIC sectors, Finance sectors \\
18 & Population density, Voter turnout, Income, Small children in childcare, CDU, SPD, Green party, FDP, AfD, 65+ year olds, Average age, Employment rate, Unemployment rate, Agriculture, forestry, fisheries, Manufacturing sector, Construction, Service sectors, TTHIC sectors, Finance sectors \\ \hline
\caption{\textbf{Regression on reaction strength of \emph{second wave} favors political variables.} For each subset size from 1 to 19 variables, all possible combinations are evaluated through exhaustive search, and the subset with the highest adjusted $R^2$ is selected.}
\label{tab:ExhSearchVariablesSecondWave}
\end{longtable}

\subsubsection{Model Selection} \label{sec:RecStrengthModelSelection}
\begin{figure}[!htp]
    \centering    \includegraphics[width=\textwidth]{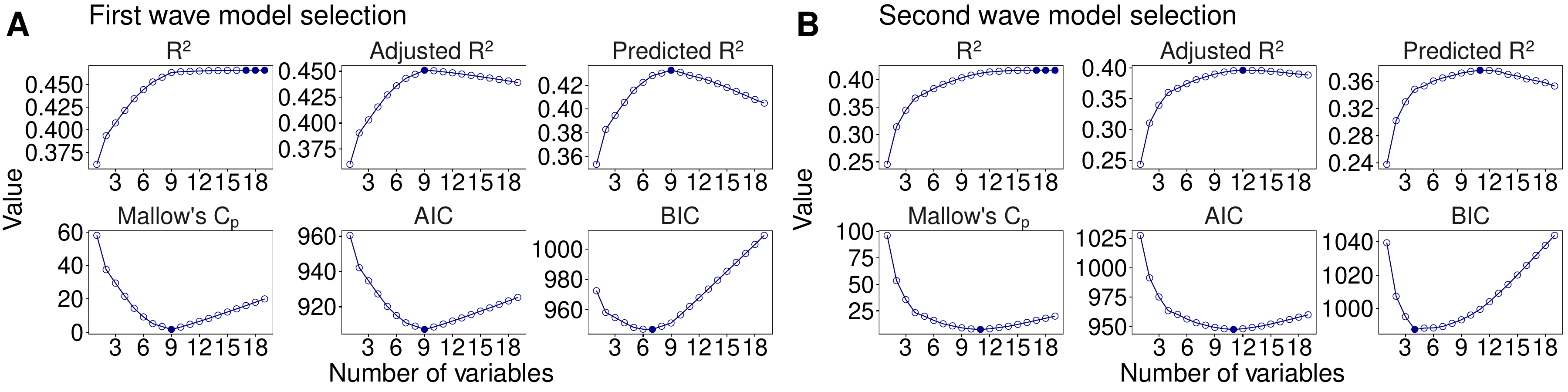}
    \caption{\textbf{Model selection criteria for the two regressions favor the 9 (first wave) and 11 (second wave) variable models.} 
    AIC is short for Akaike information criterion, BIC is short for Schwarz Bayesian information criterion. BIC favoring smaller models for both regressions is to be expected as BIC penalizes each additional parameter more strictly than AIC.
    \textbf{Left.} Model selection criteria for \emph{reaction strength first wave} as regressand. Adj. R$^2$, pred. R$^2$ Mallow's $C_p$, and AIC all favor the 9 variable model (marked with filled out circle).
    For each number of variables, the variables chosen through exhaustive search (Table~\ref{tab:ExhSearchVariablesFirstWave}) are considered. \textbf{Right.} Model selection criteria for \emph{reaction strength second wave} as regressand. Adj. R$^2$, pred. R$^2$ Mallow's $C_p$, and AIC all favor the 11 variable model (marked with filled out circle). For each number of variables, the variables chosen through exhaustive search (Table~\ref{tab:ExhSearchVariablesSecondWave}) are considered.}
    \label{fig:SpatRegExhSearchStatistics}
\end{figure}

We compare the following metrics across different subset sizes: R$^2$, adjusted R$^2$, pred R$^2$, Mallow's C$_\text{p}$, Akaike information criteria (AIC), and Schwarz Bayesian criteria (BIC).
For the first wave, the majority of selection criteria favor the nine variable model, while for the second wave, the majority of selection criteria favor the eleven variable model (Figure~\ref{fig:SpatRegExhSearchStatistics}). 
From here on forward, we consider the union of variables chosen this way: \emph{population density}, \emph{voter turnout}, \emph{income}, \emph{small children in childcare}, \emph{CDU}, \emph{SPD}, \emph{Green party}, \emph{FDP}, \emph{AfD}, \emph{average age}, \emph{unemployment rate}, \emph{employment rate}, \emph{agriculture, forestry, and fisheries}, \emph{finance sector}.
The only variable not considered in the union but favored by the model selection criteria applied to the second wave is \emph{65+ year olds}.
We exclude \emph{65+ year olds} as it strongly correlates with \emph{average age} and as we deem them to have the same effect on \emph{reaction strength}.

\FloatBarrier
\subsubsection{Regression on Reaction Strength}

We regress \emph{reaction strength first wave} and \emph{reaction strength second wave} on the union of variables chosen via the model selection of the previous paragraph.
During the first COVID-19 wave, \emph{unemployment rate} and \emph{population density} contribute most strongly and equally to explaining variance in \emph{reaction strength} (Table~\ref{tab:RegressionOutputReacStrength}).
\emph{Voter turnout} is the only other significant ($p=0.05$) explanatory variable with a positive correlation coefficient.
In contrast, \emph{agriculture, forestry, fisheries} and \emph{FDP} both contribute to explaining variance in \emph{reaction strength}, alas with a negative correlation coefficient.

\begin{table}[h]
\footnotesize
    \centering
    \sisetup{table-align-text-post=false}
    \begin{tabular}{|l|S[table-format=3.2]lS[table-format=3.3, table-space-text-post={$^{\star\star}$}]|S[table-format=3.2]lS[table-format=3.3, table-space-text-post={$^{\star\star}$}]|}
    \hline
        & \multicolumn{1}{l}{\textbf{First wave}} & & & \multicolumn{1}{l}{\textbf{Second wave}} & & \\
        \textbf{Variable} & \multicolumn{1}{l}{\textbf{Standardized}} & \textbf{Std. } & \textbf{p-value}  & \multicolumn{1}{l}{\textbf{Standardized}} & \textbf{Std. } & \textbf{p-value} \\
         & \multicolumn{1}{l}{\textbf{coefficient}} & \textbf{error} &  & \multicolumn{1}{l}{\textbf{coefficient}} & \textbf{error} & \\ \hline
        Intercept &  <0.01 & 0.04 & 1 & <0.01 & 0.04 & 1\\
        Unemployment rate & 0.30 & 0.100 & 0.002$^{\star \star}$ & 0.22 & 0.11 & 0.035$^{\star}$ \\
        Population density & 0.26 & 0.09 & 0.002$^{\star \star}$ & 0.06 & 0.07 & 0.400 \\ 
        Income & 0.19 & 0.11 & 0.069 & 0.07 & 0.10 & 0.503  \\
        Green party & 0.18 & 0.12 & 0.120  & 0.04 & 0.11 & 0.787  \\
        Finance sector & 0.11 & 0.07 & 0.131  & 0.07 & 0.09 & 0.408 \\
        Average age & 0.10 & 0.07 & 0.145 & 0.07 & 0.06 & 0.276 \\
        Voter turnout & 0.16 & 0.07 & 0.022$^{\star}$ & -0.02 & 0.07 & 0.821 \\
        Small children in childcare & 0.03 & 0.07 & 0.661 & 0.11 & 0.08 & 0.186 \\
        Employment rate & 0.03 & 0.06 & 0.624 & 0.06 & 0.05 & 0.224  \\
        SPD & 0.04 & 0.10 & 0.667  & -0.28 & 0.11 & 0.014$^{\star}$  \\
        CDU & 0.05 & 0.12 & 0.651  & -0.36 & 0.12 & 0.005$^{\star \star}$  \\ 
        Agriculture, forestry, fisheries & -0.20 & 0.05 & <0.001$^{\star \star \star}$ & -0.11 & 0.06 & 0.060 \\
        FDP & -0.14 & 0.05 & 0.003$^{\star \star}$ & -0.21 & 0.05 & <0.001$^{\star \star\star}$ \\
        AfD & 0.02 & 0.15 & 0.900 & -0.54 & 0.18 & 0.003$^{\star \star}$ \\ 
    \hline
    \textbf{Adj R.$^2$} & 0.44 & & & 0.37 & &\\
    \textbf{AIC} & 916.34 & & & 996.45 & & \\ 
    \textbf{BIC} & 980.20 & & & 1062.32 & &\\ \hline
    \end{tabular}
    \caption{\textbf{Regression on reaction strength of either wave reveals the explanatory power of population density and job-related variables during the first, and the explanatory power of political variables during the second wave.} Model outputs for regressand \emph{reaction strength first} wave (middle columns) and \emph{reaction strength second wave} (right columns). Stars indicate the significance levels with $^{\star}$ representing $p<0.05$,  $^{\star\star}$ representing $p<0.01$, $^{\star \star \star}$ representing $p<0.001$.}
    \label{tab:RegressionOutputReacStrength}
\end{table}

\FloatBarrier
\subsection{Structural Equation Modeling Explaining Cross-district Differences in Peak Incidence}
\begin{figure}[!htp]
    \centering           
    \includegraphics[width=\textwidth]{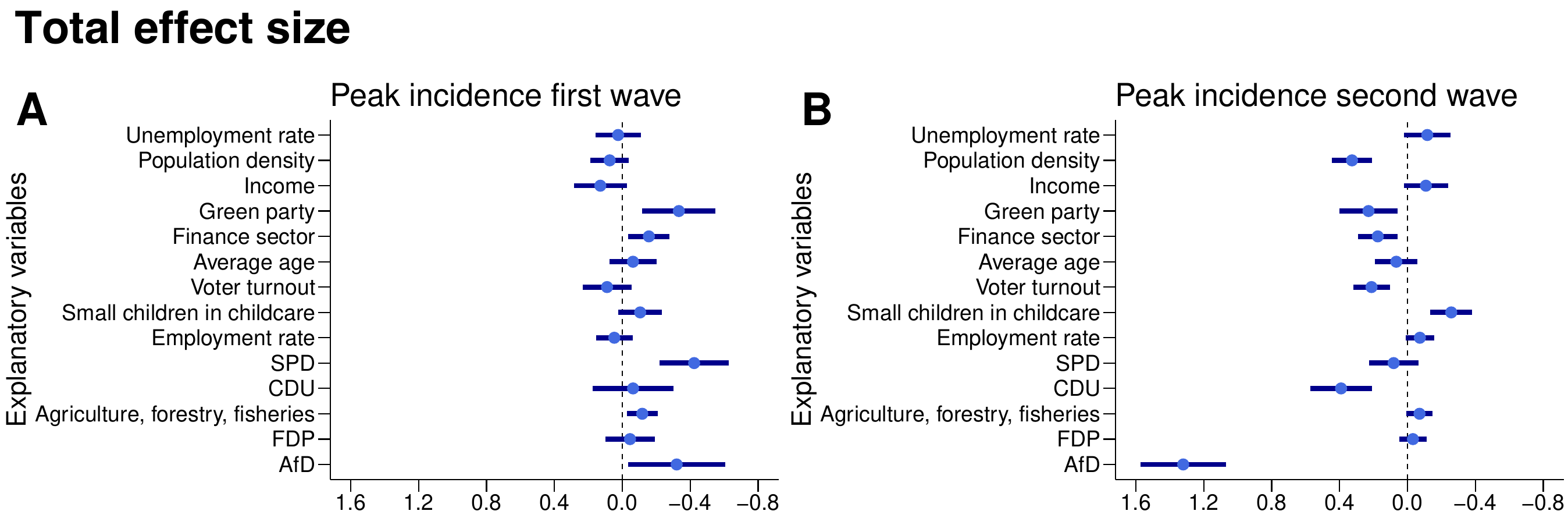}
    \caption{\textbf{Total effects of demographic, socioeconomic, and political variables on \emph{incidence}.} 
    Reaction strength serves as a mediator in the regression on incidence. Depiction of total effect size summing up direct (Fig.~\ref{fig:Regression_Contributions} C, D) and indirect (through reaction strength) effect of the socioeconomic variables on first wave (\textbf{A}) and second wave (\textbf{B}) incidence.
    }
    \label{fig:StrucEquModel}
\end{figure}

To understand which factors impact \emph{peak incidence} (here defined as the 90th percentile of the peak height of the corresponding wave), we employ a structural equation model using the R package lavaan \cite{Rosseel2012}.
We assume that \emph{reaction strength} as well as the fourteen variables chosen via model selection (Section~\ref{sec:RecStrengthModelSelection}) impact peak incidence.
However, as we have already shown that some of these fourteen socioeconomic variables also impact reaction strength, reaction strength enters the structural equation model as a mediator.
This allows the computation of the \emph{total effect size} of the fourteen economic variables, the sum of the direct effect size and the indirect effect size mediated through reaction strength (see Figure~\ref{fig:StrucEquModel} for total effect sizes and Figure~\ref{fig:Regression_Contributions} for direct effect sizes).


\FloatBarrier

\FloatBarrier
\subsection*{Data availability}

All data used in this work is publicly available (see Table~\ref{tab:DataAvailability} for details). 

\footnotesize
\begin{longtable}{|p{3.3cm}|p{9cm}|p{2.4cm}|}
\hline
\textbf{Data/Variable} & \textbf{Description} & \textbf{Source} \\ \hline
\raggedright Out-of home duration & Based on cell-based mobility data. For each zip code, the out-
of-home duration sum over all recorded persons (in seconds) is
available. Zip codes are grouped according to their district. For
each district and each day, the out-of-home duration sum (in sec-
onds) is computed, divided by the number of people living in this
district and by 3600 to obtain the out-of-home duration (in hours)
of an average person in the d. From the daily out-of-home durations, the weekly average is computed \cite{Mueller2024} & \cite{Balmer2022} \\
Temperature & Weekly average of daily maximum temperature in ($^\circ C$) for each German district. & Meteostat \cite{Meteostat2023} \\
\raggedright School vacations & School vacations for each of the 16 federal states. & Schulferien.org \cite{Schulferien2025} \\
\raggedright Public holidays & Public holidays for each of the 16 federal states. & Schulferien.org \cite{Schulferien2025} \\
\raggedright COVID-19 case numbers & COVID-19 7-Day Incidence/100.000 for each German district and all of Germany & RKI \cite{RKIIncidence2025} \\
\raggedright Population density & Extrapolation of inhabitants per km$^2$ on December~31st,~2022 based on 2011 census & Statistisches Bundesamt \cite{StatInhabitants2023} \\ 
\raggedright Income  & Primary income of private households, including private non-profit organizations in 2020, in EURO & Statistische Ämter des Bundes und der Länder \cite{StatIncome2020} \\
Unemployment rate & Unemployment rate relative to the total civilian labor force (sum of working population and registered unemployed persons) in 2020, annual average, in \% & Bundesagentur für Arbeit \cite{AgenturFuerArbeitAlq20202020} \\
Employment rate & Division of the number of employees subject to social insurance contributions on June~30th,~2020 by the population aged 15-64 on December 31 of the previous year, in \%. The 2020 \emph{employment rate} was unavailable for the following districts: Suhl, Wartburgkreis, Schmalkalden-Meiningen, Ilm-Kreis, Sonneberg, Saalfeld-Rudolstadt. For these districts, we used the 2021 \emph{employment rate} & Bundesagentur für Arbeit \cite{AgenturFuerArbeitErwerb2020} \\
Average age & Average age of the population on December~31st,~2020 & \cite{RegionalatlasAge2020}  \\
65+ year olds & Percentage of people aged 65 and older in the total population in 2020 & \cite{Deutschlandatlas65} \\
\raggedright Small children in childcare & Share of children under the age of 3 in daycare facilities/daycare centers in the age group in 2020, in \% &  Deutschlandatlas \cite{DeutschlandatlasChildren2020} \\
Voter turnout & Federal election turnout in 2021 in districts, in \% & GERDA: German Election Database \cite{Heddesheimer2025GERDA} \\ 
CDU & Share of votes for political party \emph{Christian Democratic Union of Germany} (CDU) in 2021 federal election, in \% & Regionalatlas \cite{RegionalatlasElectionResults2017} \\
SPD & Share of votes for political party \emph{Social Democratic Party of Germany} (SPD) in 2021 federal election, in \% & Regionalatlas \cite{RegionalatlasElectionResults2017} \\
Green party & Share of votes for political party \emph{Alliance 90/The Greens} (green party) in 2021 federal election, in  \% & Regionalatlas \cite{RegionalatlasElectionResults2017} \\
FDP & Share of votes for political party \emph{Free Democratic Party} (FDP) in 2021 federal election , in \% & Regionalatlas \cite{RegionalatlasElectionResults2017} \\
AfD & Share of votes for political party \emph{Alternative for Germany} (AfD) in 2021 federal election , in \% & Regionalatlas \cite{RegionalatlasElectionResults2017} \\
\raggedright Agriculture, forestry, fisheries & Proportion of employed persons in agriculture, forestry and fisheries in 2020, in \% & Regionalatlas \cite{RegionalatlasAge2020} \\
Manufacturing sector & Proportion of employed persons in the manufacturing sector in 2020, in \% & Regionalatlas \cite{RegionalatlasAge2020} \\
Construction & Proportion of employed persons working in construction in 2020, in \% & Regionalatlas \cite{RegionalatlasAge2020} \\
Service sectors & Proportion of employed persons in service sectors in 2020, in \% & Regionalatlas \cite{RegionalatlasAge2020} \\
TTHIC sectors & Proportion of employed persons in the trade, transport, hospitality, information, and communication sector in 2020, in \% & Regionalatlas \cite{RegionalatlasAge2020} \\  
Finance sector & Proportion of employed persons in financial, insurance, and business services, real estate and housing sector in 2020, in \% & Regionalatlas \cite{RegionalatlasAge2020} \\ \hline     
\caption{\textbf{Regression analysis is based solely on publicly available data from a variety of sources.} All data was available for download at the time of writing.}
\label{tab:DataAvailability}
\end{longtable}

\normalsize


\end{document}